\documentclass[floats, prd, eqnum, showpacs, nofootinbib, 
eqsecnum]{revtex4-1}

\usepackage{color,graphicx}
\usepackage{amsfonts}
\usepackage{amssymb}
\begin{document}

\title{Retrolensing by two photon spheres of a black-bounce spacetime}
\author{Naoki Tsukamoto${}^{1}$}\email{tsukamoto@rikkyo.ac.jp}
\affiliation{
${}^{1}$Department of General Science and Education, National Institute of Technology, Hachinohe College, Aomori 039-1192, Japan \\
}

\begin{abstract}
We investigate retrolensing by two photon spheres in a novel black-bounce spacetime suggested by Lobo~\textit{et al.} 
which can correspond to a Schwarzschild black hole, a regular black hole, and a traversable wormhole including an Ellis-Bronnikov wormhole.  
In a case, the wormhole has a throat which acts as a photon sphere and it has another photon sphere outside of the throat.
With the sun as a light source, an observer, and the wormhole are lined up in this order,
sunlight reflected slightly outside of the throat 
and barely outside and inside of the outer photon sphere can reach the observer.
We show that the light rays reflected by the outer photon sphere are dominant in retrolensing light curves in the case. 
\end{abstract}

\maketitle

\section{Introduction}
Compact objects such as black holes and wormholes, 
which have unstable circular light orbits~\cite{Perlick_2004_Living_Rev,Perlick:2021aok}, 
reflect light rays like a mirror due to their strong gravity
~\cite{Hagihara_1931,Darwin_1959,Atkinson_1965,Luminet_1979,Ohanian_1987,Nemiroff_1993,Frittelli_Kling_Newman_2000,Virbhadra_Ellis_2000,Bozza_Capozziello_Iovane_Scarpetta_2001,Bozza:2002zj,
Perlick:2003vg,Nandi:2006ds,Virbhadra:2008ws,Bozza_2010,Tsukamoto:2016zdu,Shaikh:2019jfr,Shaikh:2019itn,Tsukamoto:2020iez}.
This phenomena can be used as a complementary method to detect the compact objects such as the observations of  
gravitational waves~\cite{Abbott:2016blz}, shadow~\cite{Akiyama:2019cqa}, and x-ray echo~\cite{Wilkins:2021yst}.
The unstable circular light orbit and stable circular light orbit are called the photon sphere and antiphoton sphere~\footnote{An antiphoton 
sphere may cause instability of a compact object due to 
the slow decay of linear waves~\cite{Keir:2014oka,Cardoso:2014sna,Cunha:2017qtt}. The deflection angle of a light scattered by the antiphoton sphere has been studied 
in Ref.~\cite{Kudo:2022ewn}.}, respectively, 
and theoretical and observational aspects of the photon spheres and antiphoton spheres have been discussed in Refs.~\cite{Hod:2017xkz,Sanchez:1977si,Press:1971wr,Abramowicz:1990cb,Mach:2013gia,Barcelo:2000ta,Ames_1968}
and generalized and alternative surfaces of the photon spheres are also suggested~\cite{Claudel:2000yi}.

In 2002, Holz and Wheeler considered that sunlight is reflected 
by the photon sphere of a black hole passing by the Solar System~\cite{Holz:2002uf}. 
The gravitational lensing with the deflection angle of lights $\alpha \sim \pi$
in a configuration, that the light source, an observer, 
and a photon sphere as a lens are lined up in this order, is called retrolensing.
Retrolensing by photon spheres not only 
in black hole spacetimes~\cite{DePaolis:2003ad,Eiroa:2003jf,Bozza:2004kq,DePaolis:2004xe,Abdujabbarov:2017pfw,Tsukamoto:2016oca} 
but also in a wormhole spacetime~\cite{Tsukamoto:2017edq,Tsukamoto:2016zdu}
and in naked singularity spacetimes~\cite{ZamanBabar:2021zuk,Tsukamoto:2021lpm} were investigated.
The effect of light rays with the deflection angle $\alpha \sim 3\pi$ on retrolensing light curves were studied~\cite{Tsukamoto:2017edq}.

A wormhole spacetime~\cite{Visser_1995,Morris_Thorne_1988} described by general relativity 
has a structure with nontrivial topology.  
Wormholes can have photon spheres and an antiphoton sphere on and / or off their throat
and some wormholes can be black hole mimickers~\cite{Bronnikov:2005gm,Bronnikov:2006fu,Damour:2007ap,Ohgami:2015nra,Paul:2019trt,Kasuya:2021cpk,Nambu:2019sqn}. 
Simpson and Visser suggested a spacetime with a metric
\begin{eqnarray}\label{eq:11}
ds^2
&=&-\left( 1-\frac{2m}{\sqrt{r^2+a^2}} \right) dt^2+\frac{dr^2}{1-\frac{2m}{\sqrt{r^2+a^2}}}\nonumber\\
&&+(a^2+r^2)(d\vartheta^2+\sin^2\vartheta d\varphi^2), 
\end{eqnarray}
where $a$ and $m$ are nonnegative constants,
which describes a regular black hole metric and a wormhole metric with a photon sphere on one side of a region against a throat
and they called it black-bounce spacetime~\cite{Simpson:2018tsi}. 
Its metric is useful to compare black holes and wormholes. 
Its gravitational lensing~\cite{Nascimento:2020ime,Ovgun:2020yuv,Tsukamoto:2020bjm,Cheng:2021hoc},
shadows and accretion disks~\cite{Lima:2021las,Bronnikov:2021liv,Guerrero:2021ues,Bambhaniya:2021ugr,Schee:2021pdt},
the motion of S-stars around SgrA*~\cite{DellaMonica:2021fdr},
epicyclic oscillatory motion~\cite{Stuchlik:2021tcn},
field sources~\cite{Bronnikov:2021uta}, and 
a rotating counterpart~\cite{Mazza:2021rgq,Shaikh:2021yux,Islam:2021ful,Jiang:2021ajk}
have been investigated.
The alternatives of the black-bounce spacetime were also studied~\cite{Huang:2019arj,Lobo:2020ffi,Xu:2021lff,Franzin:2021vnj,Chatzifotis:2021hpg,Guo:2021wid,Barrientos:2022avi}. 

In general, a spacetime with a photon sphere and no antiphoton sphere has the common behavior of light rays near the photon sphere.
Thus, gravitational lensing in the Simpson-Visser spacetime in a strong gravitational field has a similar nature to the one in the Schwarzschild spacetime
if we consider a light source on the same side of the region against the wormhole throat. 
This is because only one photon sphere in the Simpson-Visser spacetime affects on gravitational lensing in the strong gravitational field 
under the assumed configuration. 
Recently, compact objects with two photon spheres have been 
suggested~\cite{Shaikh:2018oul,Shaikh:2019jfr,Godani:2021atr,Gan:2021pwu,Gan:2021xdl,Wang:2020emr,Wielgus:2020uqz,Guerrero:2021pxt,Peng:2021osd}
and they can be distinguished from compact objects with one photon sphere by the observations in the strong gravitational field.

Lobo~\textit{et al.}~\cite{Lobo:2020ffi} have suggested a wide class of black-bounce spacetimes with a metric 
\begin{eqnarray}
ds^2=-A(r)dt^2+\frac{dr^2}{A(r)}+(a^2+r^2)(d\vartheta^2+\sin^2\vartheta d\varphi^2), \nonumber\\
\end{eqnarray}
where $A(r)$ is given by 
\begin{eqnarray}
A(r)\equiv 1-\frac{2mr^K}{\left( a^{2N}+r^{2N} \right)^{\frac{K+1}{2N}}}.
\end{eqnarray}
If $K=0$ and $N=1$ are chosen, it recovers the metric of the Simpson-Visser spacetime~(\ref{eq:11}).
Lobo~\textit{et al.} pointed out that the metric with $K=0$ and $N\geq 2$ has a similar nature to the Simpson-Visser spacetime and 
they considered a metric function
\begin{eqnarray}\label{eq:22}
A(r)= 1-\frac{2mr^2}{\left( a^2+r^2 \right)^{3/2}}
\end{eqnarray}
by setting $K=2$ and $N=1$ to find a novel black-bounce spacetime with new properties~\cite{Lobo:2020ffi}.
It is a Schwarzschild metric for $a=0$ and $m>0$, a regular black hole metric for $0<a/m\leq 4\sqrt{3}/9$, 
a wormhole metric for $4\sqrt{3}/9<a/m$, and the Ellis-Bronnikov wormhole metric for $a>0$ and $m=0$~\cite{Ellis_1973,Chetouani_Clement_1984}. 
Tsukamoto pointed out that the metric with $K=0$ and $N= 2$ has a photon sphere off a wormhole throat and an additional photon sphere on the throat
for $4\sqrt{3}/9<a/m<2\sqrt{5}/5$~\cite{Tsukamoto:2021caq}. 
The novel black-bounce metric can be the simplest one with two photon spheres~\footnote{Note that it has 
another photon sphere in the other region against the throat. Thus, it has three photon spheres in total.}. 
Guerrero \textit{et al.} studied geometrically thin accretion disks in the black-bounce spacetime~\cite{Guerrero:2022qkh}. 

In this paper we investigate retrolensing in the novel black-bounce spacetime
by using the deflection angle of light rays in strong deflection limits~\cite{Bozza:2002zj,Tsukamoto:2016zdu,Shaikh:2019itn}~\footnote{References~\cite{Shaikh:2019jfr,Tsukamoto:2020iez,Eiroa:2002mk,Eiroa:2003jf,Bozza:2004kq,Bozza:2005tg,Bozza:2006nm,Iyer:2006cn,Bozza:2007gt,Tsukamoto:2016qro,
Ishihara:2016sfv,Tsukamoto:2016oca,Tsukamoto:2016jzh,Tsukamoto:2017edq,Hsieh:2021scb,Aldi:2016ntn,Takizawa:2021gdp,Bisnovatyi-Kogan:2022ujt} 
show the details of strong deflection limits and their applications.}. 
It would be natural to pay attention to the effects of the throat to distinguish 
between black holes and wormholes since the wormholes are characterized by the throat.
In the case of $4\sqrt{3}/9<a/m<2\sqrt{5}/5$,
the novel black-bounce spacetime can be distinguished from the Simpson-Visser spacetime
since light rays reflected slightly inside of the outer photon sphere affect retrolensing light curves in the novel black-bounce spacetime.

This paper is organized as follows. 
In Secs.~II and III, we review the deflection angle in strong deflection limits in the novel black-bounce spacetime
and we investigate the percent errors of the deflection angle in strong deflection limits, respectively.
We investigate the retrolensing in Sec.~IV and we conclude our result in Sec.~V.
In this paper, we use units in which the light speed and Newton's constant are unity.

\section{Deflection angle in strong deflection limits}
In this section, we review briefly the deflection angle of light rays in strong deflection limits~\cite{Bozza:2002zj,Tsukamoto:2016zdu,Shaikh:2019itn}
in the novel black-bounce spacetime with the function (\ref{eq:22}) suggested by Lobo~\textit{et al.}~\cite{Lobo:2020ffi}.
The spacetime is a black hole spacetime with a primary photon sphere and an event horizon for $0\leq a/m \leq 4\sqrt{3}/9$;
it is a wormhole spacetime with the primary photon sphere off a throat 
and a secondary photon sphere on the throat for $4\sqrt{3}/9 < a/m < 2\sqrt{5}/5$, 
and it is a wormhole spacetime with the primary photon sphere on the throat for $2\sqrt{5}/5<a/m$.
Here and hereinafter, we call the photon spheres in order from the largest: 
We call the outer (inner) photon sphere primary (secondary) photon sphere for $4\sqrt{3}/9 < a/m < 2\sqrt{5}/5$
and we call only one photon sphere primary photon sphere in other cases.  

We consider that light rays come from spatial infinity, they are reflected by a compact object, and they go into spatial infinity. 
The deflection angle of the light ray with an impact parameter $b$ is given by~\cite{Tsukamoto:2021caq}
\begin{eqnarray}\label{eq:def}
\alpha=2\int^\infty_{r_0} \frac{dr}{\sqrt{a^2+r^2} \sqrt{\frac{a^2+r^2}{b^2}-A}}-\pi,
\end{eqnarray}
where $r_0$ is the closest distance of the light ray.
The deflection angle of the light ray reflected little outside of the primary photon sphere 
in a strong deflection limit $b\rightarrow b_{\mathrm{m}}+0$, 
where $b_{\mathrm{m}}$ is a critical impact parameter,
is expressed by 
\begin{eqnarray}\label{eq:defab}
\alpha
&=&-\bar{a} \log \left( \frac{b}{b_{\mathrm{m}}} -1 \right) + \bar{b} \nonumber\\
&&+O\left( \left( \frac{b}{b_{\mathrm{m}}} -1 \right) \log \left( \frac{b}{b_{\mathrm{m}}} -1 \right) \right),
\end{eqnarray}
where $\bar{a}$ and $\bar{b}$ are parameters determined 
by $a$ and $m$~\cite{Bozza:2002zj,Tsukamoto:2016jzh,Shaikh:2019jfr,Tsukamoto:2021caq}\footnote{The order of the vanishing term is estimated as in Ref.~\cite{Bozza:2002zj} 
as $O\left( b/b_{\mathrm{m}} -1 \right)$ 
while it should be read as $O\left( \left( b/b_{\mathrm{m}} -1 \right) \log \left( b/b_{\mathrm{m}} -1 \right) \right)$ as shown in Refs.~\cite{Iyer:2006cn,Tsukamoto:2016qro,Tsukamoto:2016jzh}}.
On the other hand, the deflection angle of light rays reflected slightly inside the primary photon sphere in a strong deflection limit $b\rightarrow b_{\mathrm{m}}-0$ 
is given by 
\begin{eqnarray}\label{eq:defcd}
\alpha
&=&-\bar{c} \log \left( \frac{b_{\mathrm{m}}}{b} -1 \right) + \bar{d} \nonumber\\
&&+O\left( \left( \frac{b_{\mathrm{m}}}{b} -1 \right) \log \left( \frac{b_{\mathrm{m}}}{b} -1 \right) \right),
\end{eqnarray}
where $\bar{c}$ and $\bar{d}$ are obtained by $a$ and $m$~\cite{Shaikh:2019itn,Tsukamoto:2021caq}\footnote{Notice that the deflection 
angle in the strong deflection limit $b\rightarrow b_{\mathrm{m}}-0$ in Refs.~\cite{Shaikh:2019jfr,Tsukamoto:2021lpm} is given by
\begin{eqnarray}\label{eq:ap}
\alpha
&=&-\bar{c} \log \left( \frac{b^2_{\mathrm{m}}}{b^2} -1 \right) + \bar{d}^\prime \nonumber\\
&&+O\left( \left( \frac{b_{\mathrm{m}}}{b} -1 \right) \log \left( \frac{b_{\mathrm{m}}}{b} -1 \right) \right).
\end{eqnarray}
In this paper, however, we use Eq.~(\ref{eq:defcd}) according to Ref.~\cite{Tsukamoto:2021caq}.
From an approximation 
\begin{eqnarray}
\left( \frac{b^2_{\mathrm{m}}}{b^2} -1 \right) \sim 2 \left( \frac{b_{\mathrm{m}}}{b} -1 \right),
\end{eqnarray}
we obtain a relation 
\begin{eqnarray}
\bar{d}=-\bar{c}\log 2 + \bar{d}^\prime.
\end{eqnarray}
The difference between $\bar{d}$ and $\bar{d}^\prime$ causes apparent differences 
between the formulas of retrolensing in the following section and the ones in Refs.~\cite{Shaikh:2019jfr,Tsukamoto:2021lpm}.}.  
The details of the following calculations of the deflection angle in the strong deflection limits are shown in Ref.~\cite{Tsukamoto:2021caq}.

\subsection{For $a/m\leq 4\sqrt{3}/9$}
For $a/m\leq 4\sqrt{3}/9$, light rays with the impact parameter $b\rightarrow b_{\mathrm{m}}+0$, 
where the critical impact parameter $b_{\mathrm{m}}$ is given by
\begin{eqnarray}\label{eq:bm}
b_{\mathrm{m}}=\frac{\left(r_{\mathrm{m}}^2+a^2\right)^\frac{5}{4}}{\left( r_{\mathrm{m}}^2-2a^2 \right)^\frac{1}{2} m^\frac{1}{2}},
\end{eqnarray}
are reflected slightly outside of the photon sphere at $r=r_{\mathrm{m}}$, where $r_{\mathrm{m}}$ 
is the largest solution of an equation 
\begin{eqnarray}\label{eq:rm}
(2a^2-3r^2)m+(a^2+r^2)^\frac{3}{2}=0.
\end{eqnarray}
The parameters $\bar{a}$ and $\bar{b}$ in the deflection angle~(\ref{eq:defab}) are obtained as
\begin{eqnarray}\label{eq:bara}
\bar{a}= \frac{\left( a^2+r_{\mathrm{m}}^2 \right)^\frac{5}{4}}{r_{\mathrm{m}} \sqrt{3(r_{\mathrm{m}}^2-4a^2)m}}
\end{eqnarray}
and 
\begin{equation}\label{eq:barb}
\bar{b}= \bar{a} \log \frac{6r_{\mathrm{m}}^4(r_{\mathrm{m}}^2-4a^2)}{(a^2+r_{\mathrm{m}}^2)^2(r_{\mathrm{m}}^2-2a^2)}+I_\mathrm{R}-\pi,
\end{equation}
respectively, where $I_\mathrm{R}$ is given by
\begin{eqnarray}
I_{\mathrm{R}}=\int^1_0 g(z) dz, 
\end{eqnarray}
and $g(z)$ is defined as
\begin{eqnarray}\label{eq:gz}
g(z) 
\equiv && \frac{2r_\mathrm{m}}{\sqrt{r_\mathrm{m}^2+a^2(1-z)^2}}
 \left[ \frac{r_\mathrm{m}^2+a^2(1-z)^2}{b_\mathrm{m}^2} \right. \nonumber\\
&&\left.-(1-z)^2+\frac{2mr_\mathrm{m}^2(1-z)^3}{\left[ r_\mathrm{m}^2+a^2(1-z)^2 \right]^\frac{3}{2}} \right] ^{-\frac{1}{2}} \nonumber\\
&&-\frac{2\left( a^2+r_{\mathrm{m}}^2 \right)^\frac{5}{4}}{r_\mathrm{m}\sqrt{3m\left( r_{\mathrm{m}}^2-4a^2 \right)} \left| z \right|}.
\end{eqnarray}

\subsection{For $4\sqrt{3}/9<a/m<2\sqrt{5}/5$}
In the case of $4\sqrt{3}/9<a/m\leq 2\sqrt{5}/5$,
the wormhole has a primary photon sphere off a throat at $r=r_{\mathrm{m}}$ which is the larger solution of Eq.~(\ref{eq:rm}) 
and the secondary photon sphere on the throat at $r=r_{\mathrm{sc}}=0$.

\subsubsection{Rays reflected barely outside of the primary photon sphere}
Light rays reflected slightly outside of the primary photon sphere are calculated by using
parameters $\bar{a}$, $\bar{b}$ in the deflection angle~(\ref{eq:defab}), and the critical impact parameter $b_{\mathrm{m}}$ 
given by Eq.~(\ref{eq:bara}), Eq.~(\ref{eq:barb}), and Eq.~(\ref{eq:bm}), respectively.

\subsubsection{Rays reflected little inside the primary photon sphere}
We obtain parameters $\bar{c}$ and $\bar{d}$ in the deflection angle~(\ref{eq:defcd}) of
rays reflected slightly inside of the primary photon sphere as
\begin{equation}
\bar{c}\equiv  \frac{2\left( a^2+r_{\mathrm{m}}^2 \right)^\frac{5}{4}}{r_{\mathrm{m}} \sqrt{3(r_{\mathrm{m}}^2-4a^2)m}}
\end{equation}
and
\begin{equation}
\bar{d}=\bar{c} \log \left(  \frac{6r_\mathrm{m}^4(r_\mathrm{m}^2-4a^2)}{(r_\mathrm{m}^2+a^2)^2 (r_\mathrm{m}^2-2a^2)} \left( \frac{r_\mathrm{m}}{r_\mathrm{c}}-1 \right) \right) +I_\mathrm{r}-\pi,
\end{equation}
where $I_\mathrm{r}$ is defined by 
\begin{equation}
I_\mathrm{r}\equiv \int^1_{1-\frac{r_\mathrm{m}}{r_\mathrm{c}}} g(z) dz, 
\end{equation}
where $r_\mathrm{c}$ is the smaller positive solution of $V(r)=0$.
Here, $V(r)$ is the effective potential of the critical impact parameter $b=b_\mathrm{m}$ defined by
\begin{equation}
V(r)\equiv \left\{ \left[ \frac{1}{a^2+r^2}-\frac{2mr^2}{(a^2+r^2)^\frac{5}{2}} \right] b^2 -1 \right\} E^2,
\end{equation}
where $E$ is the conserved energy of the light ray, 
and $g(z)$ and $b_\mathrm{m}$ are given by Eq.~(\ref{eq:gz}) and Eq.~(\ref{eq:bm}), respectively.

\subsubsection{Rays reflected barely outside of the secondary photon sphere}
The deflection angle of light rays reflected little outside of the secondary photon sphere at the throat~$r=r_{\mathrm{sc}}=0$ 
in a strong deflection limit $b\rightarrow b_{\mathrm{sc}}+0$, 
where $b_{\mathrm{sc}}=a$ is the critical impact parameter for the secondary photon sphere,
is written in 
\begin{eqnarray}\label{eq:alsec}
\alpha
&=&-\bar{a} \log \left( \frac{b}{b_{\mathrm{sc}}} -1 \right) + \bar{b} \nonumber\\
&&+O\left( \left( \frac{b}{b_{\mathrm{sc}}} -1 \right) \log \left( \frac{b}{b_{\mathrm{sc}}} -1 \right) \right),
\end{eqnarray}
where $\bar{a}$ and $\bar{b}$ are 
are obtained as
\begin{equation}\label{eq:bara2}
\bar{a}=\sqrt{\frac{a}{a+2m}}
\end{equation}
and
\begin{equation}\label{eq:barb2}
\bar{b}= \bar{a} \log \frac{4(a+2m)}{a}+I_\mathcal{R}-\pi, 
\end{equation}
respectively, where $I_\mathcal{R}$ is defined by
\begin{eqnarray}
I_\mathcal{R}&\equiv&\int^1_0 \left[ \frac{2}{z(2-z)}\sqrt{\frac{a}{a+2m(1-z)^3}} \right. \nonumber\\
&& \left. -\frac{1}{z}\sqrt{\frac{a}{a+2m}} \right] dz.
\end{eqnarray}

\subsection{For $a/m>2\sqrt{5}/5$}
For $a/m>2\sqrt{5}/5$, rays with $b\rightarrow b_{\mathrm{m}}+0$, where the critical impact parameter is $b_{\mathrm{m}}=a$, 
are reflected slightly outside of the primary photon sphere on the throat $r=r_{\mathrm{m}}=0$, 
with $\bar{a}$ and $\bar{b}$ in the deflection angle~(\ref{eq:defab}) given by Eq.~(\ref{eq:bara2}) and Eq.~(\ref{eq:barb2}), respectively.

\section{Percent errors of the deflection angle in the strong deflection limits}
We define the percent errors of the deflection angles $\alpha$ of Eqs. (\ref{eq:defab}), (\ref{eq:defcd}), and (\ref{eq:alsec}) in the strong deflection limits
against the deflection angle of Eq. (\ref{eq:def}) as 
\begin{eqnarray}\label{eq:errorab}
\frac{\alpha \: \mathrm{of\: Eq.}\: (\ref{eq:defab})-\alpha \: \mathrm{of\: Eq.}\: (\ref{eq:def}) }{\alpha \: \mathrm{of\: Eq.}\: (\ref{eq:def})} \times 100,
\end{eqnarray}
\begin{eqnarray}\label{eq:errorcd}
\frac{\alpha \: \mathrm{of\: Eq.}\: (\ref{eq:defcd})-\alpha \: \mathrm{of\: Eq.}\: (\ref{eq:def}) }{\alpha \: \mathrm{of\: Eq.}\: (\ref{eq:def})} \times 100,
\end{eqnarray}
and
\begin{eqnarray}\label{eq:erroras2}
\frac{\alpha \: \mathrm{of\: Eq.}\: (\ref{eq:alsec})-\alpha \: \mathrm{of\: Eq.}\: (\ref{eq:def}) }{\alpha \: \mathrm{of\: Eq.}\: (\ref{eq:def})} \times 100,
\end{eqnarray}
respectively.
We plot the percent errors (\ref{eq:errorab}), (\ref{eq:errorcd}), and (\ref{eq:erroras2}) against the deflection angle of Eq. (\ref{eq:def}) 
in Fig. 1, 2, and 3, respectively.
\begin{figure}[htbp]
\begin{center}
\includegraphics[width=80mm]{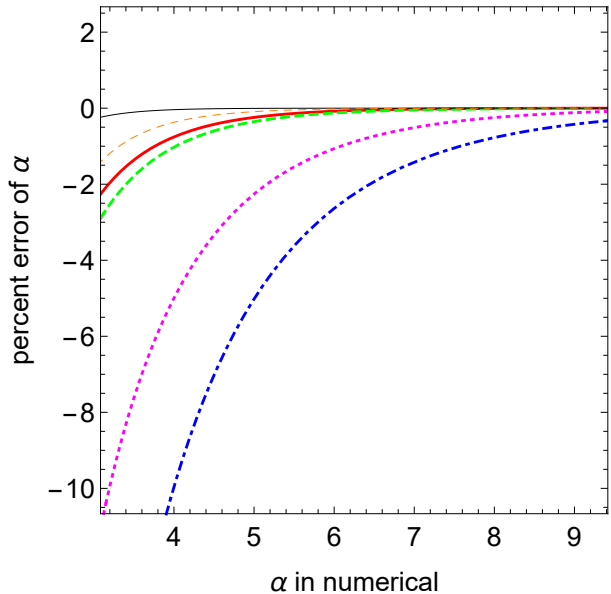}
\end{center}
\caption{
The percent error (\ref{eq:errorab}) of the deflection angle $\alpha$ of a light ray reflected nearly outside of the primary photon sphere against $\alpha$ of Eq.~(\ref{eq:def}).
Thick solid (red), thick dashed (green), thick dotted (magenta), thick dot-dashed (blue), 
thin solid (black), and thin dashed (orange) curves denote the percent error (\ref{eq:errorab}) for $a/m=0$, $0.5$, $0.83$, $0.86$, $0.9$, and $3.0$, respectively.
}
\label{err1}
\end{figure}
\begin{figure}[htbp]
\begin{center}
\includegraphics[width=80mm]{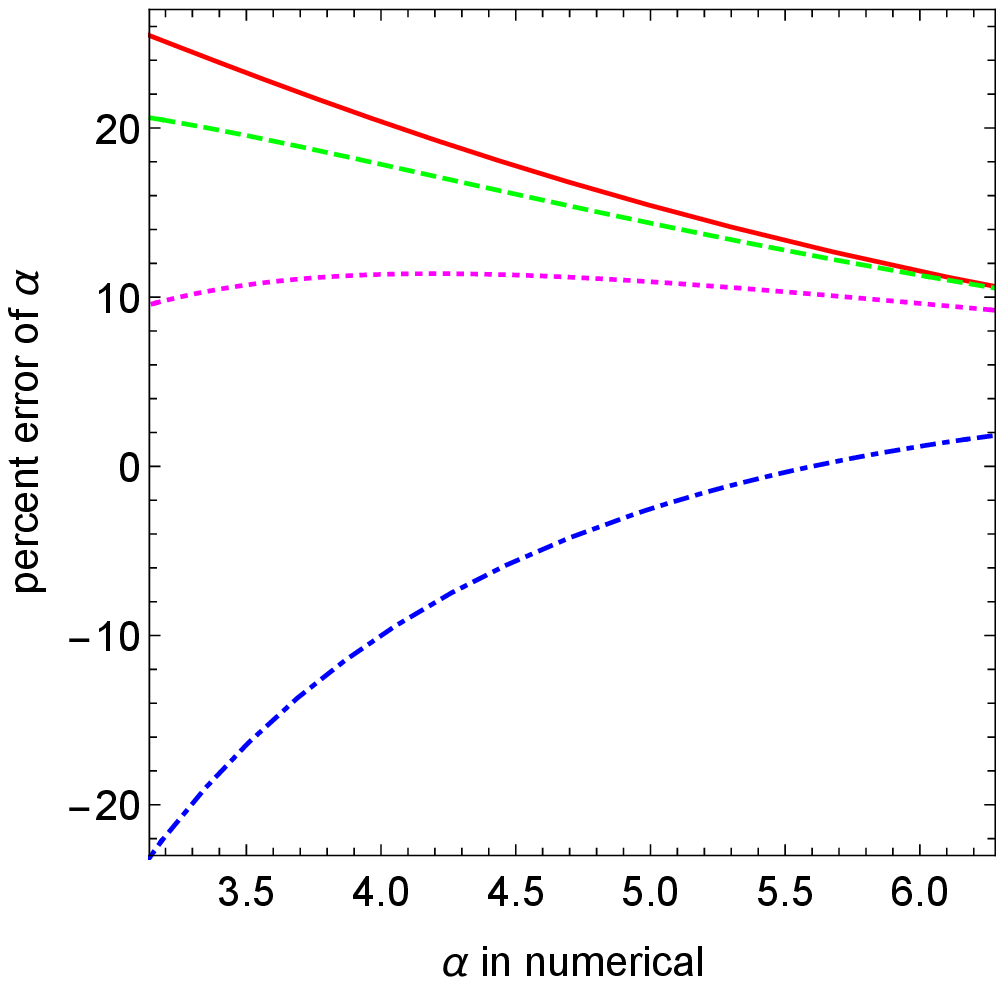}
\end{center}
\caption{
The percent error (\ref{eq:errorcd}) of the deflection angle $\alpha$ of a light ray reflected slightly inside of the primary photon sphere against $\alpha$ of Eq.~(\ref{eq:def}).
Solid (red), dashed (green), dotted (magenta), and dot-dashed (blue), 
curves denote the percent error (\ref{eq:errorcd}) for $a/m=0.77$, $0.8$, $0.83$, and $0.86$, respectively.
}
\label{err2}
\end{figure}
\begin{figure}[htbp]
\begin{center}
\includegraphics[width=80mm]{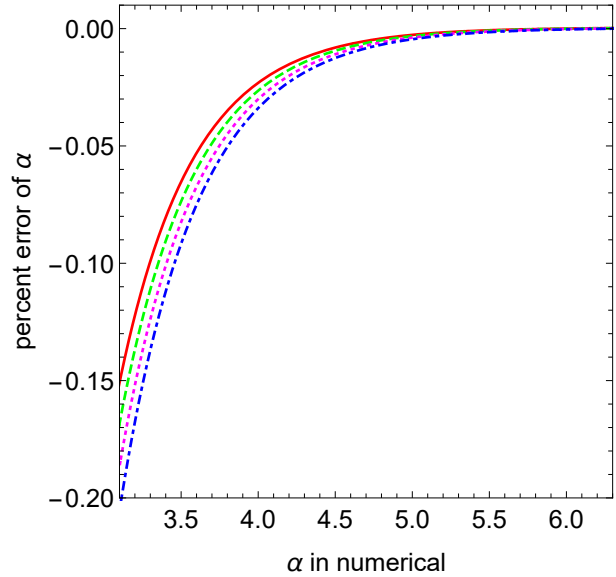}
\end{center}
\caption{
The percent error (\ref{eq:erroras2}) of the deflection angle $\alpha$ of a light ray reflected nearly outside of the secondary photon sphere against $\alpha$ of Eq.~(\ref{eq:def}).
Solid (red), dashed (green), dotted (magenta), and dot-dashed (blue), 
curves denote the percent error (\ref{eq:erroras2}) for $a/m=0.77$, $0.8$, $0.83$, and $0.86$, respectively.
}
\label{err3}
\end{figure}
Figure 1 shows that the absolute value of the percent error of retrolensing with $\alpha \sim \pi$ by the Schwarzschild black hole is about $2\%$ 
and that it is larger than gravitational lensing by the photon sphere of the black hole in usual lens configurations with $\alpha \sim 2\pi$.
From Figs.~1 and 3, we notice that the absolute value of the percent errors 
of light rays reflected slightly outside of the primary and secondary photon spheres on the wormhole throat can be small even if we consider retrolensing.  
From Fig.~2, the absolute value of the percent error of the light rays reflected barely inside of the primary photon sphere is relatively large.

The absolute value of the percent error is large in an almost marginally unstable photon sphere case $a/m = 2\sqrt{5}/5 + \epsilon$, where $0<\epsilon \ll 1$, while 
it is small in a case with $a/m = 2\sqrt{5}/5 - \epsilon$ as shown in Figs.~1-3. 
This is caused by the degeneracy of the outer photon sphere and an antiphoton sphere into a marginally unstable photon sphere 
in a limit $a/m \rightarrow 2\sqrt{5}/5-0$. 
The deflection angle of the light rays reflected by the marginally unstable photon sphere would diverge nonlogarithmically in the strong deflection limits.

\section{Retrolensing}
In this section, we investigate retrolensing by the photon spheres.

\subsection{Lens equation}
A light ray emitted by the sun S is reflected by the photon sphere L with the deflection angle $\alpha$, 
reaches to an observer O,
and the observer sees it as an image I as shown in Fig.~4. 
We use the Ohanian lens equation~\cite{Ohanian_1987,Bozza:2004kq,Bozza:2008ev} expressed by
\begin{equation}\label{eq:lens}
\beta=\pi-\bar{\alpha}(\theta)+\theta+\bar{\theta},
\end{equation}
where $\beta\equiv \angle$OLS is a source angle, 
\begin{equation}
\bar{\alpha} \equiv \alpha \qquad  ( \mathrm{mod}\;  2\pi ),
\end{equation}
is an effective deflection angle,
$\theta\equiv \angle$IOL is an image angle, and 
$\bar{\theta}$ is an angle between the ray and a line LS.
\begin{figure}[htbp]
\begin{center}
\includegraphics[width=80mm]{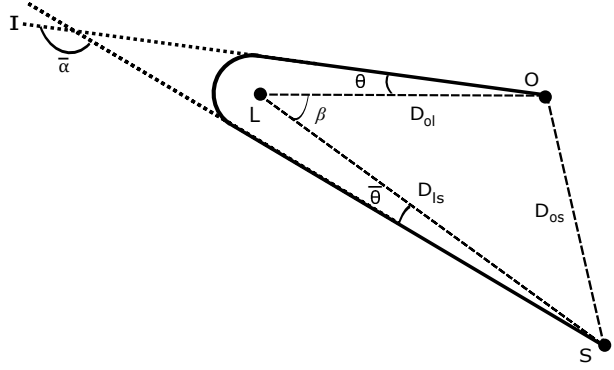}
\end{center}
\caption{Lens configuration. A light ray emitted by the sun S with a source angle $\beta \equiv \angle$OLS 
is reflected with a deflection angle $\alpha$ by a lens object L 
and it is observed by an observer O as an image I with an image angle $\theta$.
The effective deflection angle $\bar{\alpha}$ is given by $\bar{\alpha}= \alpha -2\pi n$, where $n$ is the winding number of the ray,
$\bar{\theta}$ is an angle between the ray and the line LS, 
and $D_\mathrm{ol}$, $D_\mathrm{ls}$, and $D_\mathrm{os}$ are distances between O and L, between L and S, and between O and S, respectively. 
}
\label{configuration}
\end{figure}

We assume that L, O, and S are almost aligned in this order. 
Under the assumption, we obtain $\beta \sim 0$,  
$\bar{\alpha} \sim \pi$,
$\alpha \sim \pi +2\pi n$, and 
$D_\mathrm{ls}=D_\mathrm{ol}+D_\mathrm{os}$,
where $n$ is the winding number of the light, and $D_\mathrm{ls}$, $D_\mathrm{ol}$, and $D_\mathrm{os}$
are distances between L and S, between O and L, and between O and S, respectively.
We also assume that $\theta=b_{\mathrm{m}}/D_\mathrm{ol}\ll 1$ and $\bar{\theta}=b_{\mathrm{m}}/D_\mathrm{ls}\ll 1$ 
and we neglect the terms of the small angles $\theta$ and $\bar{\theta}$ in the Ohanian lens equation. 
We regard the sun as a uniform-luminous disk with a finite size on the observer's sky.

\subsection{Image separations and magnifications}

\subsubsection{Light rays reflected slightly outside of the primary and secondary photon spheres}
We consider the image separations and magnifications of rays reflected little outside of the photon spheres 
in the retrolensing configuration. First, we concentrate on the primary photon sphere.
Note that we use $\bar{a}$~(\ref{eq:bara}) and $\bar{b}$~(\ref{eq:barb}) in this case. 
By using Eq.~(\ref{eq:defab}), $\alpha=\bar{\alpha}+2\pi n$, and $b=\theta D_\mathrm{ol}$, and the assumptions, we obtain the positive solution of 
the lens equation~(\ref{eq:lens}) as
\begin{equation}\label{eq:theta+10}
\theta=\theta_{n}^{\mathrm{out}}(\beta)\equiv \left( 1+e^{\left[ \bar{b}-(1+2n)\pi+\beta  \right] /\bar{a}   } \right) \theta_\mathrm{m},
\end{equation}
where $\theta_\mathrm{m}$ is the image angle of the primary photon sphere given by $\theta_\mathrm{m}\equiv b_\mathrm{m}/D_\mathrm{ol}$, 
and it is plotted in Fig.~5.
\begin{figure}[htbp]
\begin{center}
\includegraphics[width=80mm]{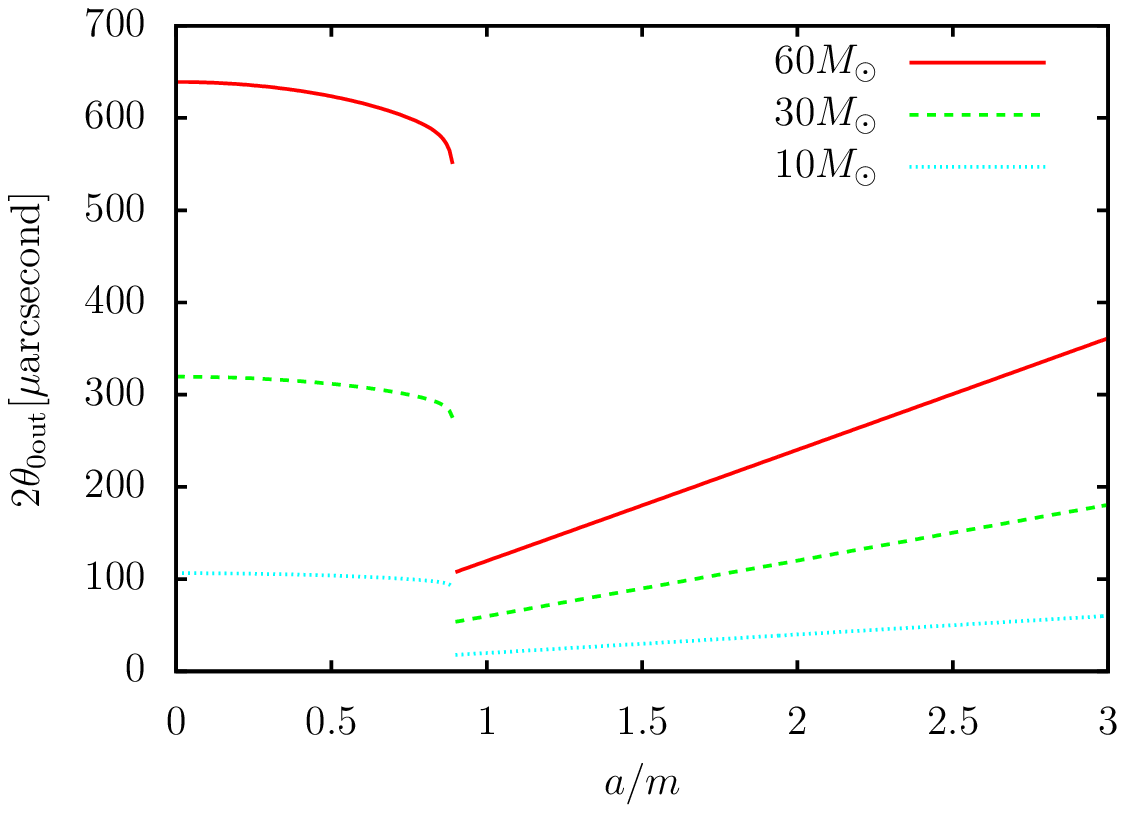}\\
\includegraphics[width=80mm]{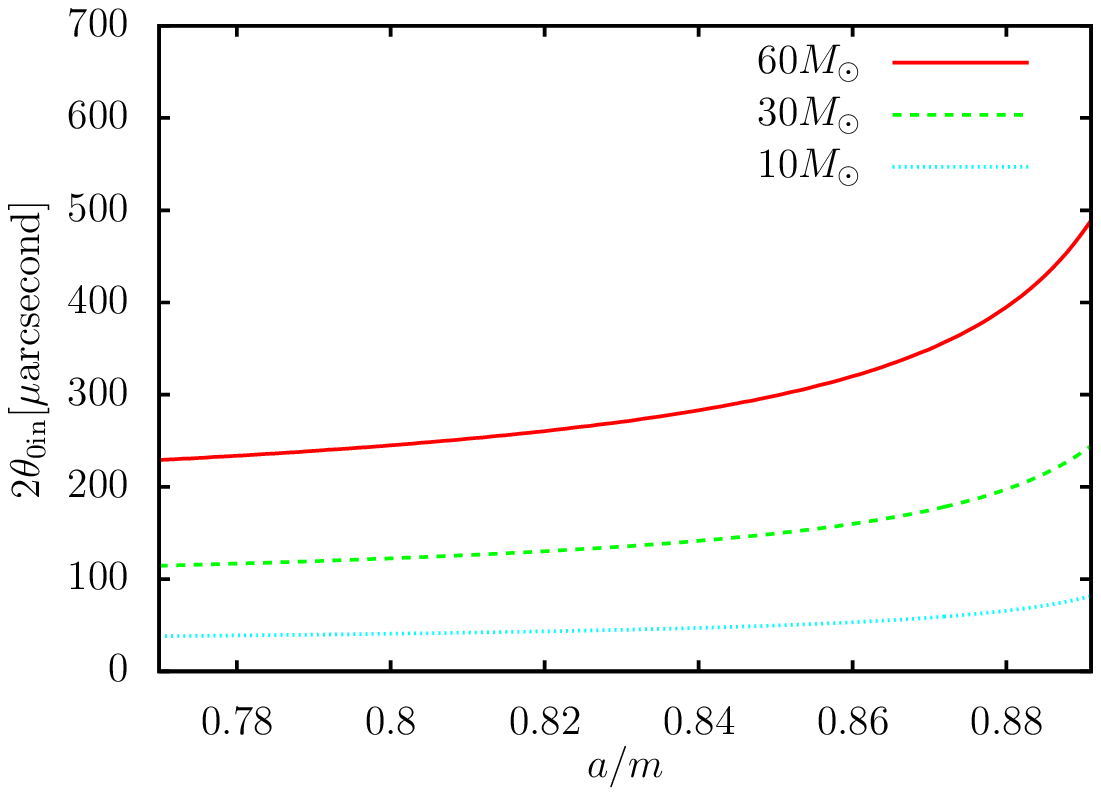}\\
\includegraphics[width=80mm]{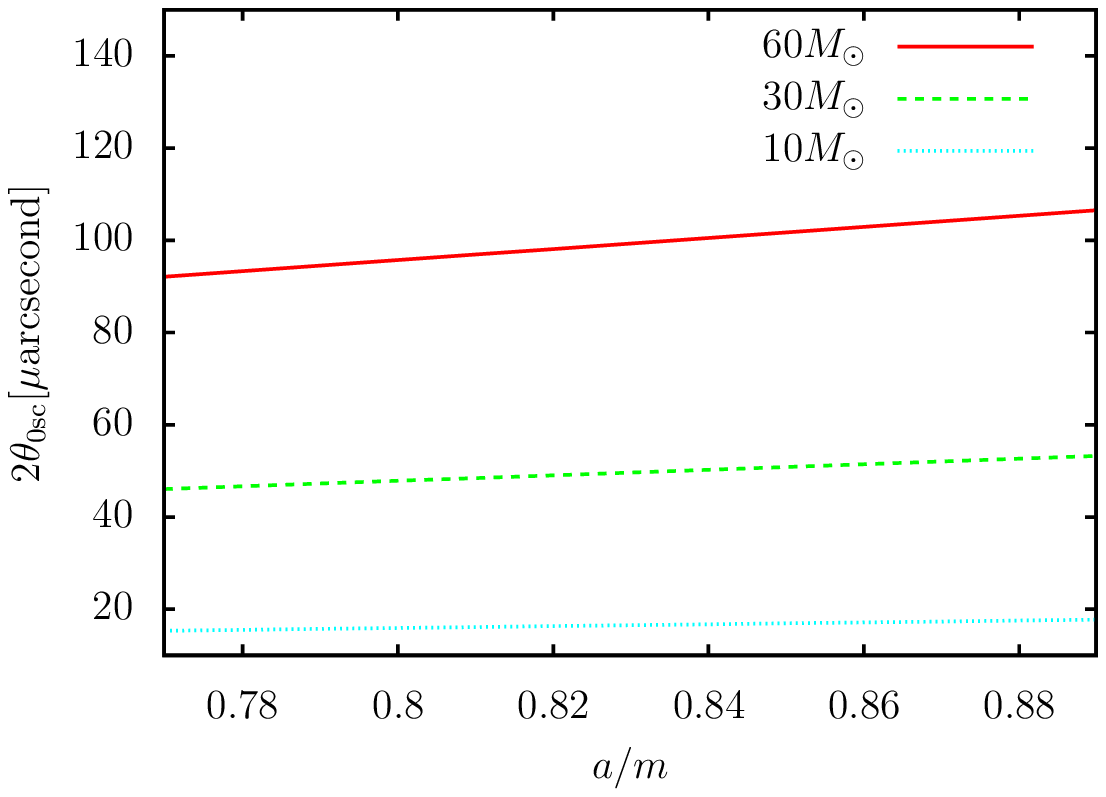}
\end{center}
\caption{Image separations $2\theta_{0}^{\mathrm{out}}$ (top), $2\theta_{0}^{\mathrm{in}}$ (middle), 
and $2\theta_{0}^{\mathrm{sc}}$ (bottom) with $D_\mathrm{ol}=0.01$ pc, 
$\beta=0$, and $n=0$ as functions of $a/m$. Solid (red), dashed (green), and dotted (cyan) curves denote 
the image separations for $m=60M_{\odot}$, $30M_{\odot}$, and $10M_{\odot}$, respectively.
}
\label{image}
\end{figure}
Its magnification is obtained as
\begin{eqnarray}\label{eq:magnification}
\mu_{n}^{\mathrm{out}}(\beta)
&=&-\frac{D_\mathrm{os}^{2}}{D_\mathrm{ls}^{2}} s(\beta)\theta_{n}^{\mathrm{out}}\frac{d\theta_{n}^{\mathrm{out}}}{d\beta} \nonumber\\
&=&-\frac{D_\mathrm{os}^{2}}{D_\mathrm{ls}^{2}}\frac{\theta_\mathrm{m}^{2}e^{\left[\bar{b}-(1+2n)\pi\right]/\bar{a}}}{\bar{a}} \nonumber\\
&&\times \left( 1+e^{\left[\bar{b}-(1+2n)\pi\right]/\bar{a}} \right) s(\beta).
\end{eqnarray}
where $s(\beta)$ is an integral over the disk on a source plane~\cite{Witt:1994,Nemiroff:1994uz,Alcock:1997fi} given by
\begin{equation}
s(\beta)
= \frac{1}{\pi \beta_\mathrm{s}^{2}} \int_{\mathrm{disk}} d\beta' d\phi,
\end{equation}
where $\beta_\mathrm{s}\equiv R_\mathrm{s}/D_\mathrm{ls}$ is the dimensionless radius of the sun, where $R_\mathrm{s}$ is the radius of the sun, 
$\beta'$ is a dimensionless radial coordinate normalized by $D_\mathrm{ls}$ on the source plane, 
and $\phi$ is an azimuthal coordinate around an origin which is an intersection point between an axis $\beta=0$ and 
the source plane~\footnote{Note that $s(\beta)=1/\beta$ for a point source.}. 
The function $s(\beta)$ is expressed by
\begin{eqnarray}
s(\beta)
&=& \frac{2}{\pi \beta_\mathrm{s}^{2}} \left[ \pi(\beta_\mathrm{s}-\beta) \right. \nonumber\\
&&\left.+\int^{\beta+\beta_\mathrm{s}}_{-\beta+\beta_\mathrm{s}} \arccos  \frac{\beta^{2}+\beta'^{2}-\beta^{2}_\mathrm{s}}{2\beta \beta'} d\beta' \right]
\end{eqnarray}
and 
\begin{eqnarray}
s(\beta)
=\frac{2}{\pi \beta_\mathrm{s}^{2}} \int^{\beta+\beta_\mathrm{s}}_{\beta-\beta_\mathrm{s}} \arccos \frac{\beta^{2}+\beta'^{2}-\beta^{2}_\mathrm{s}}{2\beta \beta'} d\beta' 
\end{eqnarray}
for $\beta \leq \beta_\mathrm{s}$ and $\beta_\mathrm{s} < \beta$, respectively.
Note that we get $s(0)=2/\beta_\mathrm{s}$ for the perfectly aligned case.
There are negative solutions $\theta \sim -\theta_{n}^{\mathrm{out}}(\beta)$ for each winding number $n$ 
and their magnifications are obtained as $-\mu_{n}^{\mathrm{out}}$ approximately.
The total magnification of the couple of images for all $n$ is given by
\begin{eqnarray}
\mu_\mathrm{tot}^{\mathrm{out}}(\beta)
&\equiv& 2 \sum_{n=0}^{\infty} \left| \mu_{n}^{\mathrm{out}}(\beta) \right|  \nonumber\\
&=& 2\frac{D_\mathrm{os}^{2}}{D_\mathrm{ls}^{2}}
\frac{\theta_\mathrm{m}^{2}}{\bar{a}}\left|s(\beta)\right| 
\left[ \frac{e^{(\bar{b}-\pi)/\bar{a}}}{1-e^{-2\pi/\bar{a}}}+\frac{e^{2(\bar{b}-\pi)/\bar{a}}}{1-e^{-4\pi/\bar{a}}} \right] \nonumber\\
\end{eqnarray}
and, in the perfectly aligned case, it becomes 
\begin{equation}\label{eq:aligned_magnification}
\mu_\mathrm{tot}^{\mathrm{out}}(0)
=4\frac{D_\mathrm{os}^{2}}{D_\mathrm{ls}^{2}}\frac{\theta_\mathrm{m}^{2}}{\bar{a}\beta_\mathrm{s}}
\left[ \frac{e^{(\bar{b}-\pi)/\bar{a}}}{1-e^{-2\pi/\bar{a}}}+\frac{e^{2(\bar{b}-\pi)/\bar{a}}}{1-e^{-4\pi/\bar{a}}} \right].
\end{equation}

Second, we comment on the secondary photon sphere. 
In this case.
if we read $\theta_\mathrm{m}$ as $\theta^\mathrm{sc}\equiv b_\mathrm{sc}/D_\mathrm{ol}$
and if we use $\bar{a}$ and $\bar{b}$ given by Eqs.~(\ref{eq:bara2}) and (\ref{eq:barb2}), respectively,
the above formulas for the images of the rays reflected little outside of the primary photon sphere can be used as the ones of the secondary photon sphere.

\subsubsection{Light rays reflected barely inside of the primary photon sphere}
By using the deflection angle~(\ref{eq:defcd}), the image angle of rays reflected slightly inside of the primary photon sphere 
is obtained as the positive solution of the lens equation: 
\begin{equation}\label{eq:theta01}
\theta=\theta_{n}^{\mathrm{in}}(\beta)\equiv 
\frac{\theta_{\mathrm{m}}}{1+ e^{ \left[\bar{d}-(1+2n)\pi+\beta \right] /\bar{c} }}.
\end{equation}
Its magnification for each $n$ is obtained as
\begin{equation}
\mu_{n}^{\mathrm{in}}(\beta)
=\frac{D_\mathrm{os}^{2}}{D_\mathrm{ls}^{2}}
\frac{\theta_\mathrm{m}^{2}e^{\left[ \bar{d}-(1+2n)\pi \right]/\bar{c}}}{\bar{c}\left( 1+e^{\left[ \bar{d}-(1+2n)\pi \right] /\bar{c}} \right)^3}s(\beta).
\end{equation}
Due to a negative solution $\theta \sim -\theta_{n}^{\mathrm{in}}(\beta)$ for each winding number~$n$,
the total magnification of the couple of images for all the winding number $n$ is given by
\begin{eqnarray}
\mu_{\mathrm{tot}}^{\mathrm{in}}(\beta)
&=& 2 \sum_{n=0}^{\infty} \left|  \mu_{n}^{\mathrm{in}}(\beta) \right| \nonumber\\
&=& 2\sum_{n=0}^{\infty} 
\frac{D_\mathrm{os}^{2}}{D_\mathrm{ls}^{2}}
\frac{\theta_\mathrm{m}^{2}e^{\left[ \bar{d}-(1+2n)\pi \right]/\bar{c}}}{\bar{c}\left( 1+e^{\left[ \bar{d}-(1+2n)\pi \right] / \bar{c}} \right)^3}
\left|s(\beta)\right| \nonumber\\
\end{eqnarray}
and it is, in the perfect-aligned case, 
\begin{equation}
\mu_{\mathrm{tot}}^{\mathrm{in}}(0)
=4\sum_{n=0}^{\infty} \frac{D_\mathrm{os}^{2}}{D_\mathrm{ls}^{2}}
\frac{\theta_\mathrm{m}^{2}e^{ \left[ \bar{d}-(1+2n)\pi \right]/\bar{c}}}{\bar{c}\left( 1+e^{\left[ \bar{d}-(1+2n)\pi \right]/\bar{c}} \right)^3\beta_\mathrm{s}}.
\end{equation}

The apparent magnification of the retrolensing in the perfect-aligned case is shown in Fig.~6. 
If there are light rays reflected barely inside of the primary photon sphere, they are dominant.
On the other hand, if there are rays reflected slightly outside of the secondary photon sphere, 
their effect on light curves can be ignored since they are fainter than the others. 
\begin{figure}[htbp]
\begin{center}
\includegraphics[width=80mm]{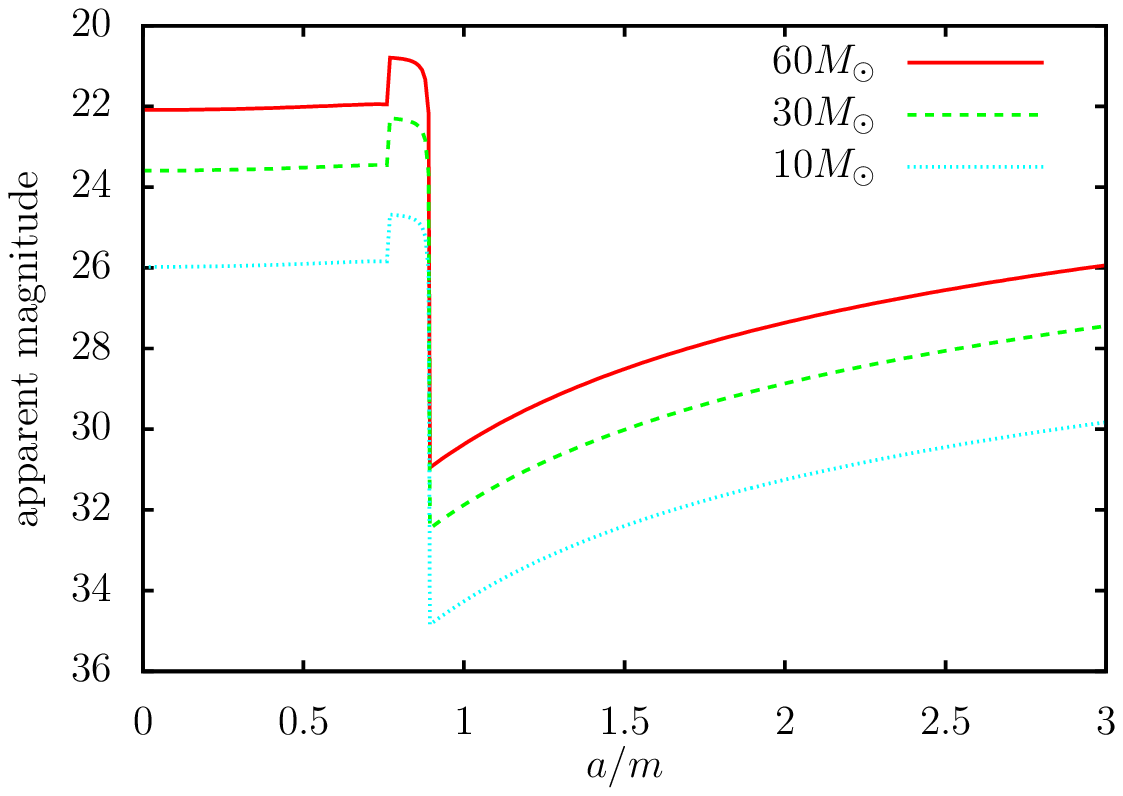}
\end{center}
\caption{The apparent magnification with $\beta=0$ and $D_\mathrm{ol}=0.01$pc for given $a/m$.
Solid (red), dashed (green), and dotted (cyan) curves denote the apparent magnification of the retrolensing caused by the photon spheres with 
the mass $m=60M_\odot$, $30M_\odot$, and $10M_\odot$, respectively.}
\end{figure}

\subsection{Retrolensing light curves}
We assume that the lens L is at rest against the observer O as shown in Fig.~7.
\begin{figure}[htbp]
\begin{center}
\includegraphics[width=80mm]{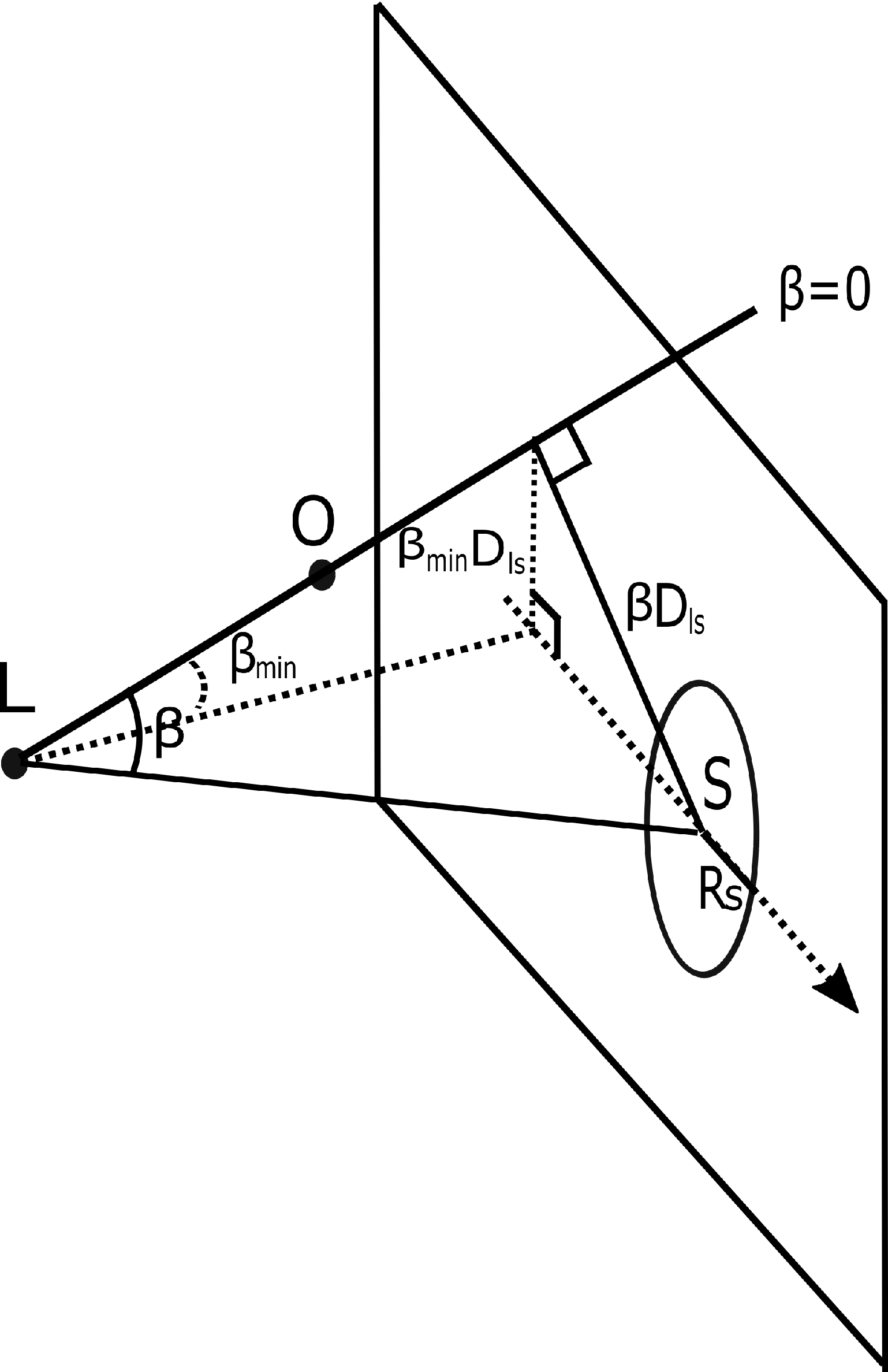}
\end{center}
\caption{Motion of the sun S with a radius $R_\mathrm{s}$ on a source plane which is orthogonal to an optical axis $\beta=0$. 
We assume that an observer O and a lens L are at rest and the sun moves with the orbital velocity $v=30$km/s on the source plane. 
The smallest source angle is denoted by $\beta_\mathrm{min}$.
}
\end{figure}
The retrolensing light curves by the photon spheres are shown in Fig.~8.
\begin{figure}[htbp]
\begin{center}
\includegraphics[width=80mm]{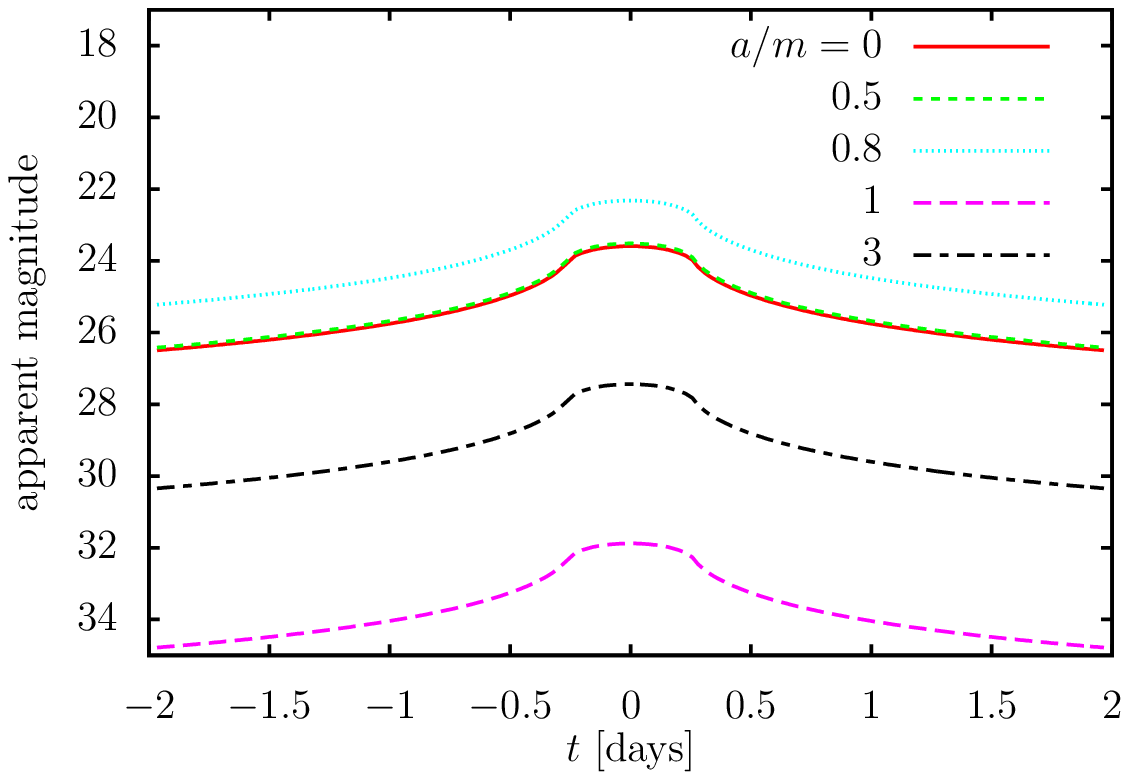}\\
\includegraphics[width=80mm]{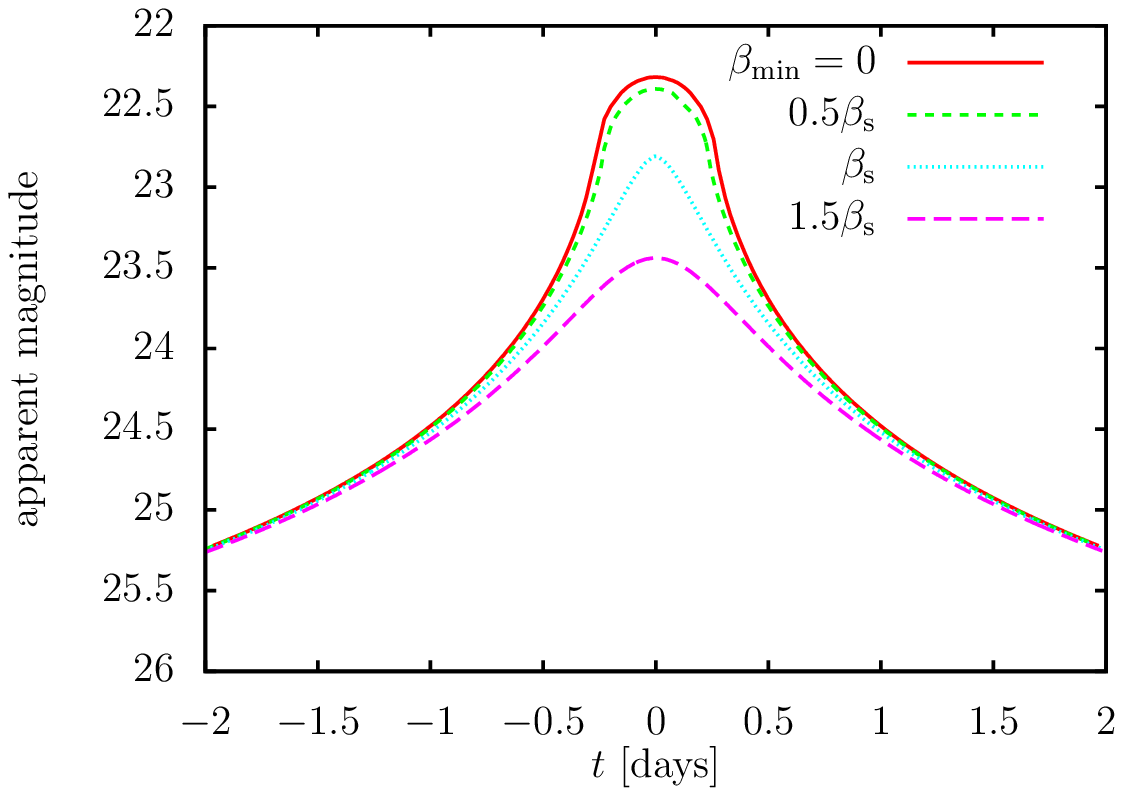}\\
\includegraphics[width=80mm]{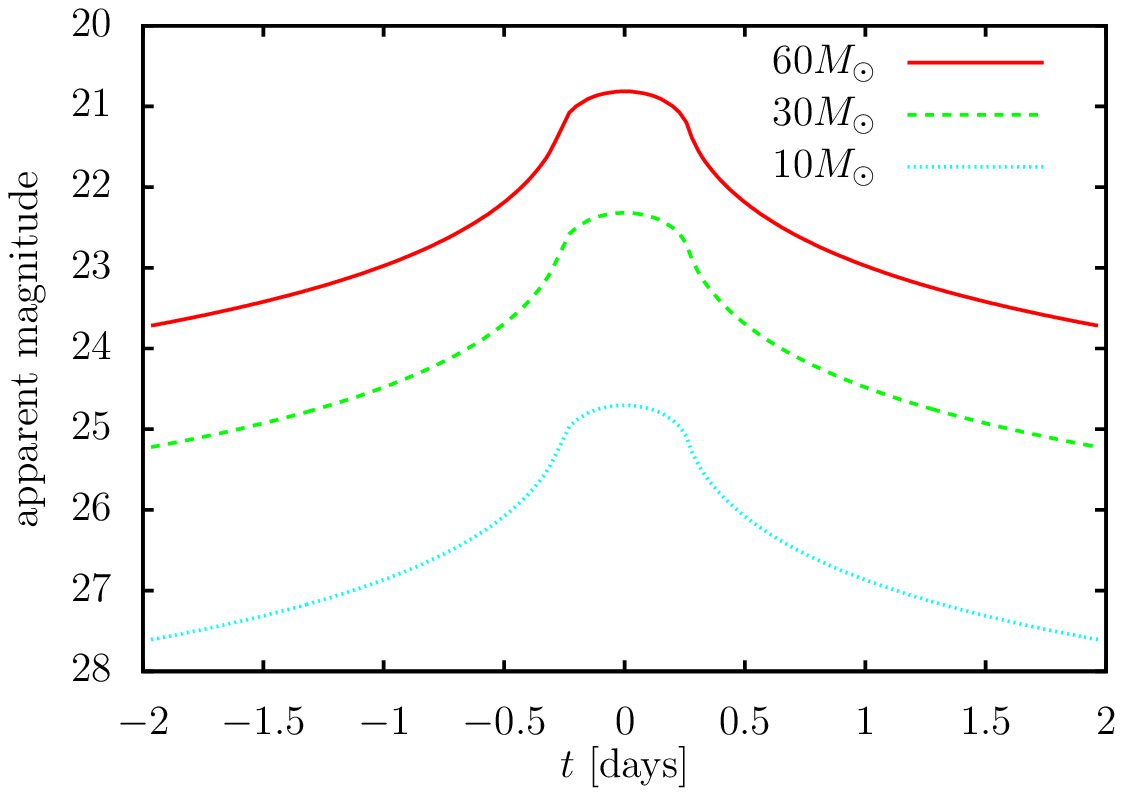}
\end{center}
\caption{Light curves with $D_\mathrm{ol}=0.01$pc. 
Top: solid (red), dashed (green), dotted (cyan), long-dashed (magenta), and long-dashed-short-dashed (black) curves 
denote light curves with $a/m=0$, $0.5$, $0.8$, $1$, and $3$, respectively, 
and $m=30 M_\odot$ and $\beta_\mathrm{min}=0$. 
Middle: solid (red), dashed (green), dotted (cyan), and long-dashed (magenta) curves denote 
light curves with $\beta_\mathrm{min}=0$, $0.5\beta_\mathrm{s}$, $\beta_\mathrm{s}$, and $1.5\beta_\mathrm{s}$, respectively, 
and $m=30 M_\odot$ and $a/m=0.8$.  
Bottom: solid (red), dashed (green), and dotted (cyan) curves denote 
light curves with $m=60 M_\odot$, $30 M_\odot$, and $10 M_\odot$, respectively, 
and $\beta_\mathrm{min}=0$ and $a/m=0.8$.  
}
\end{figure}

\section{conclusion} 
We have investigated retrolensing in a novel black-bounce spacetime~\cite{Lobo:2020ffi}, which is a black hole spacetime with a photon sphere 
or a wormhole spacetime with one or two photon spheres on one side of a region against a throat. 
If the wormhole has two photon spheres there, 
light rays reflected by the inner photon sphere on a throat can be ignored since the magnifications are dimmer
than the ones by the outer photon sphere.
However, the retrolensing by the wormhole with two photon spheres is brighter than the black hole with one photon sphere 
since rays reflected slightly not only outside but also inside of the outer photon sphere of the wormhole reach an observer 
due to absence of an event horizon.
We can distinguish the retrolensing from other variable phenomena 
since the retrolensing light curve has characteristic shapes as shown in Fig.~8;  
it can be observed on the ecliptic, 
and it has precise solar spectra~\cite{Holz:2002uf}. 

The light rays reflected near the outer photon sphere with the winding number $n=1$ reach the observer a few milliseconds later than ones with $n=0$. 
However, the rays with $n\geq 1$ are fainter than the lights with $n=0$. 
Therefore, we can ignore the effect of the rays with $n\geq 1$ on the retrolensing light curves.  

We have shown the percent errors of the deflection angles of retrolensing in strong deflection limits.   
We have found that the percent errors depend on the parameters of the metric and 
the cases of reflections: (i) a reflection nearly outside of the photon sphere on the throat, 
(ii) a reflection barely outside of the photon sphere off the throat, 
(iii) a reflection slightly inside of the photon sphere off the throat.
We have shown that the case (iii) has larger error than the cases (i) and (ii). 
Thus, we should improve approximations in the strong deflection limit for the retrolensing of the rays reflected nearly inside of the photon sphere. 
One may use an improved approximation (\ref{eq:ap}) which was applied in an usual lens configuration~\cite{Shaikh:2019itn,Tsukamoto:2021caq} 
and in the retrolensing configuration~\cite{Tsukamoto:2021caq}.  
We will investigate the details of the improved approximation in a forthcoming paper~\cite{Tsukamoto}. 

We do not consider the case of $a/m=2\sqrt{5}/5$ because the deflection angles of the light rays reflected 
by the outer photon do not diverge logarithmically in the forms of Eqs.~(\ref{eq:defab}) and (\ref{eq:defcd}). 
Note that the observables of the outer photon sphere with the vanishing winding number $n=0$ 
obtained by using Eqs. (\ref{eq:defab}) and (\ref{eq:defcd}) in the strong deflection limits have relativity large error for $a/m \sim 2\sqrt{5}/5-\epsilon$.
In the case of $a/m = 2\sqrt{5}/5$, the outer photon sphere and an antiphoton sphere degenerate into a marginally unstable photon sphere 
and the deflection angles of the light rays reflected by it would diverge nonlogarithmically in the strong deflection limits.
We can obtain the deflection angle of the light rays reflected slightly outside of the marginally unstable photon sphere in the strong deflection limit
as well as Ref.~\cite{Tsukamoto:2020iez}. On the other hand, 
the case of the rays reflected slightly inside of the marginally unstable photon sphere 
and retrolensing by the marginally unstable photon sphere are left as future works.

\section*{Acknowledgements}
The author thanks an anonymous referee for his or her useful comments.
%
%\appendix
%\section{Ricci scalar}


\begin{thebibliography}{99}

\bibitem{Perlick_2004_Living_Rev}
V. Perlick,
Living Rev. Relativity {\bf7}, 9 (2004).

\bibitem{Perlick:2021aok}
V.~Perlick and O.~Y.~Tsupko,
%``Calculating black hole shadows: Review of analytical studies,''
Phys. Rept. {\bf 947}, 1 (2022).
%[arXiv:2105.07101 [gr-qc]].

\bibitem{Hagihara_1931} 
Y.~Hagihara, 
%Theory of the relativistic trajectories in a gravitational field of Schwarzschild$B!I(B,
Jpn.\ J.\ Astron.\ Geophys., {\bf 8}, 67 (1931).

\bibitem{Darwin_1959}
C. Darwin,
Proc. R. Soc. Lond. A {\bf 249}, 180 (1959).

\bibitem{Atkinson_1965}
R.~d'~E. Atkinson,
%"On light tracks near a very massive star",
Astron. J. {\bf 70}, 517 (1965).

\bibitem{Luminet_1979}
J.-P. Luminet,  
Astron. Astrophys. {\bf 75}, 228 (1979).

\bibitem{Ohanian_1987}
H. C. Ohanian, 
Am. J. Phys. {\bf 55}, 428 (1987).

\bibitem{Nemiroff_1993}
R. J. Nemiroff,  
Am. J. Phys. {\bf 61}, 619 (1993).

\bibitem{Frittelli_Kling_Newman_2000}
S. Frittelli, T. P. Kling, and E. T. Newman,
Phys. Rev. D {\bf 61}, 064021 (2000).

\bibitem{Virbhadra_Ellis_2000}
K. S. Virbhadra and G. F. R. Ellis,
Phys. Rev. D {\bf 62}, 084003 (2000).

\bibitem{Bozza_Capozziello_Iovane_Scarpetta_2001}
V. Bozza, S. Capozziello, G. Iovane, and G. Scarpetta,
Gen. Relativ. Gravit. {\bf 33}, 1535 (2001).

\bibitem{Bozza:2002zj} 
  V.~Bozza,
  %``Gravitational lensing in the strong field limit,''
  Phys.\ Rev.\ D {\bf 66}, 103001 (2002).
%  [gr-qc/0208075].

\bibitem{Perlick:2003vg}
V.~Perlick,
%``On the Exact gravitational lens equation in spherically symmetric and static space-times,''
Phys. Rev. D {\bf 69}, 064017 (2004).
%[arXiv:gr-qc/0307072 [gr-qc]].

\bibitem{Nandi:2006ds}
K.~K.~Nandi, Y.~Z.~Zhang, and A.~V.~Zakharov,
%``Gravitational lensing by wormholes,''
Phys. Rev. D {\bf 74}, 024020 (2006).
%[arXiv:gr-qc/0602062 [gr-qc]].

\bibitem{Virbhadra:2008ws}
K.~S.~Virbhadra,
%``Relativistic images of Schwarzschild black hole lensing,''
Phys. Rev. D {\bf 79}, 083004 (2009).
%[arXiv:0810.2109 [gr-qc]].

\bibitem{Bozza_2010}
V. Bozza,
Gen. Relativ. Gravit. {\bf 42}, 2269 (2010).

\bibitem{Tsukamoto:2016zdu}
  N.~Tsukamoto and T.~Harada,
  %``Light curves of light rays passing through a wormhole,''
  Phys.\ Rev.\ D {\bf 95}, 024030 (2017).
  %  [arXiv:1607.01120 [gr-qc]].

\bibitem{Shaikh:2019jfr} 
  R.~Shaikh, P.~Banerjee, S.~Paul, and T.~Sarkar,
  %``Strong gravitational lensing by wormholes,''
  JCAP {\bf 07}, 028 (2019).
%  [arXiv:1905.06932 [gr-qc]].

\bibitem{Shaikh:2019itn} 
  R.~Shaikh, P.~Banerjee, S.~Paul, and T.~Sarkar,
  %``Analytical approach to strong gravitational lensing from ultracompact objects,''
  Phys.\ Rev.\ D {\bf 99}, 104040 (2019).
%  [arXiv:1903.08211 [gr-qc]].

\bibitem{Tsukamoto:2020iez}
N.~Tsukamoto,
%``Deflection angle of a light ray reflected by a general marginally unstable photon sphere in a strong deflection limit,''
Phys. Rev. D {\bf 102}, 104029 (2020).
%[arXiv:2008.12244 [gr-qc]].

\bibitem{Abbott:2016blz}
  B.~P.~Abbott {\it et al.} [LIGO Scientific and Virgo Collaborations],
  %``Observation of Gravitational Waves from a Binary Black Hole Merger,''
  Phys.\ Rev.\ Lett.\  {\bf 116}, 061102 (2016).
%  [arXiv:1602.03837 [gr-qc]].
  
\bibitem{Akiyama:2019cqa} 
  K.~Akiyama {\it et al.} [Event Horizon Telescope Collaboration],
  %``First M87 Event Horizon Telescope Results. I. The Shadow of the Supermassive Black Hole,''
  Astrophys.\ J.\  {\bf 875}, L1 (2019).

\bibitem{Wilkins:2021yst}
D.~R.~Wilkins, L.~C.~Gallo, E.~Costantini, W.~N.~Brandt and R.~D.~Blandford,
%``Light bending and X-ray echoes from behind a supermassive black hole,''
Nature \textbf{595}, no.7869, 657-660 (2021).
%[arXiv:2107.13555 [astro-ph.HE]].

\bibitem{Keir:2014oka} 
  J.~Keir,
  %``Slowly decaying waves on spherically symmetric spacetimes and ultracompact neutron stars,''
  Class.\ Quant.\ Grav.\  {\bf 33}, 135009 (2016).
%  [arXiv:1404.7036 [gr-qc]].

\bibitem{Cardoso:2014sna} 
  V.~Cardoso, L.~C.~B.~Crispino, C.~F.~B.~Macedo, H.~Okawa, and P.~Pani,
  %``Light rings as observational evidence for event horizons: long-lived modes, ergoregions and nonlinear instabilities of ultracompact objects,''
  Phys.\ Rev.\ D {\bf 90}, 044069 (2014).
%  [arXiv:1406.5510 [gr-qc]].

\bibitem{Cunha:2017qtt} 
  P.~V.~P.~Cunha, E.~Berti, and C.~A.~R.~Herdeiro,
  %``Light-Ring Stability for Ultracompact Objects,''
  Phys.\ Rev.\ Lett.\  {\bf 119}, 251102 (2017).
%  [arXiv:1708.04211 [gr-qc]].

\bibitem{Kudo:2022ewn}
R.~Kudo and H.~Asada,
%``Nondivergent deflection of light around a photon sphere of a compact object,''
[arXiv:2201.01946 [gr-qc]].

\bibitem{Hod:2017xkz} 
  S.~Hod,
  %``Upper bound on the radii of black-hole photonspheres,''
  Phys.\ Lett.\ B {\bf 727}, 345 (2013);
%  [arXiv:1701.06587 [gr-qc]].
%\bibitem{Hod:2017zpi} 
  S.~Hod,
  %``On the number of light rings in curved spacetimes of ultra-compact objects,''
  Phys.\ Lett.\ B {\bf 776}, 1 (2018).
%  [arXiv:1710.00836 [gr-qc]].

\bibitem{Sanchez:1977si} 
  N.~G.~Sanchez,
  %``Absorption and Emission Spectra of a Schwarzschild Black Hole,''
  Phys.\ Rev.\ D {\bf 18}, 1030 (1978);
%\bibitem{Decanini:2010fz} 
  Y.~Decanini, A.~Folacci, and B.~Raffaelli,
  %``Unstable circular null geodesics of static spherically symmetric black holes, Regge poles and quasinormal frequencies,''
  Phys.\ Rev.\ D {\bf 81}, 104039 (2010);
%  [arXiv:1002.0121 [gr-qc]].
%\bibitem{Wei:2011zw} 
  S.~W.~Wei, Y.~X.~Liu, and H.~Guo,
  %``Relationship between High-Energy Absorption Cross Section and Strong Gravitational Lensing for Black Hole,''
  Phys.\ Rev.\ D {\bf 84}, 041501 (2011).
%  [arXiv:1103.3822 [hep-th]].

  \bibitem{Press:1971wr} 
  W.~H.~Press,
  %``Long Wave Trains of Gravitational Waves from a Vibrating Black Hole,''
  Astrophys.\ J.\  {\bf 170}, L105 (1971);
%\bibitem{Goebel_1972} 
 C. J. Goebel, 
 %Comments on the $B!H(Bvibrations$B!I(B of a black hole,
Astrophys.\ J.\ {\bf 172}, L95 (1972);
%\bibitem{Stefanov:2010xz} 
  I.~Z.~Stefanov, S.~S.~Yazadjiev, and G.~G.~Gyulchev,
  %``Connection between Black-Hole Quasinormal Modes and Lensing in the Strong Deflection Limit,''
  Phys.\ Rev.\ Lett.\  {\bf 104}, 251103 (2010);
%  [arXiv:1003.1609 [gr-qc]].
%\bibitem{Raffaelli:2014ola} 
  B.~Raffaelli,
  %``Strong gravitational lensing and black hole quasinormal modes: Towards a semiclassical unified description,''
  Gen.\ Rel.\ Grav.\  {\bf 48},  16 (2016).
%  [arXiv:1412.7333 [gr-qc]].

\bibitem{Abramowicz:1990cb} 
  M.~A.~Abramowicz,
  %``Centrifugal Force: A Few Surprises,''
Mon.\ Not.\ Roy.\ Astr.\ Soc.\ {\bf 245}, 733 (1990);
%\bibitem{Hasse_Perlick_2002}
W. Hasse and V. Perlick,
Gen. Relativ. Gravit. {\bf 34}, 415 (2002).

\bibitem{Mach:2013gia} 
  P.~Mach, E.~Malec, and J.~Karkowski,
  %``Spherical steady accretion flows: Dependence on the cosmological constant, exact isothermal solutions, and applications to cosmology,''
  Phys.\ Rev.\ D {\bf 88}, 084056 (2013);
%  [arXiv:1309.1252 [gr-qc]].
%\bibitem{Chaverra:2015bya} 
  E.~Chaverra and O.~Sarbach,
  %``Radial accretion flows on static spherically symmetric black holes,''
  Class.\ Quant.\ Grav.\  {\bf 32}, 155006 (2015);
%  [arXiv:1501.01641 [gr-qc]].
%\bibitem{Cvetic:2016bxi} 
  M.~Cvetic, G.~W.~Gibbons, and C.~N.~Pope,
  %``Photon Spheres and Sonic Horizons in Black Holes from Supergravity and Other Theories,''
  Phys.\ Rev.\ D {\bf 94}, 106005 (2016);
%  [arXiv:1608.02202 [gr-qc]].
% \bibitem{Koga:2016jjq} 
  Y.~Koga and T.~Harada,
  %``Correspondence between sonic points of ideal photon gas accretion and photon spheres,''
  Phys.\ Rev.\ D {\bf 94}, 044053 (2016).
%  [arXiv:1601.07290 [gr-qc]].
 
\bibitem{Barcelo:2000ta}
C.~Barcelo and M.~Visser,
%``Brane surgery: Energy conditions, traversable wormholes, and voids,''
Nucl. Phys. B {\bf 584}, 415 (2000);
%[arXiv:hep-th/0004022 [hep-th]].
%\bibitem{Koga:2020gqd}
Y.~Koga,
%``Photon surfaces as pure tension shells: Uniqueness of thin shell wormholes,''
Phys. Rev. D {\bf 101}, 104022 (2020).
%[arXiv:2003.10859 [gr-qc]].

\bibitem{Ames_1968} 
W. L. Ames and K. S. Thorne, 
%``The optical appearance of a star that is collapsing through its gravitational radius,''
Astrophys.\ J. {\bf 151}, 659 (1968);
%\bibitem{Synge:1966okc} 
  J.~L.~Synge,
  %``The Escape of Photons from Gravitationally Intense Stars,''
  Mon.\ Not.\ Roy.\ Astron.\ Soc.\  {\bf 131}, 463 (1966);
%\bibitem{Yoshino:2019qsh} 
  H.~Yoshino, K.~Takahashi, and K.~i.~Nakao,
  %``How does a collapsing star look?,''
  Phys.\ Rev.\ D {\bf 100}, 084062 (2019).
%  [arXiv:1908.04223 [gr-qc]].

\bibitem{Claudel:2000yi} 
  C.~M.~Claudel, K.~S.~Virbhadra, and  G.~F.~R.~Ellis,
  %``The Geometry of photon surfaces,''
  J.\ Math.\ Phys.\  {\bf 42}, 818 (2001);
%\bibitem{Gibbons:2016isj}
G.~W.~Gibbons and C.~M.~Warnick,
%``Aspherical Photon and Anti-Photon Surfaces,''
Phys. Lett. B {\bf 763}, 169 (2016);
%[arXiv:1609.01673 [gr-qc]].
%\bibitem{Yoshino:2017gqv}
H.~Yoshino, K.~Izumi, T.~Shiromizu, and  Y.~Tomikawa,
%``Extension of photon surfaces and their area: Static and stationary spacetimes,''
PTEP {\bf 2017}, 063E01 (2017);
%[arXiv:1704.04637 [gr-qc]].
%\bibitem{Galtsov:2019fzq}
D.~V.~Gal'tsov and K.~V.~Kobialko,
%``Photon trapping in static axially symmetric spacetime,''
Phys. Rev. D {\bf 100}, 104005 (2019);
%[arXiv:1906.12065 [gr-qc]].
%\bibitem{Siino:2019vxh}
M.~Siino,
%``Generalization of Photon Sphere by Causal Structure--- In Order to See Dynamical Black Hole Shadow,''
Class. Quantum Grav. {\bf 38}, 025005 (2021);
%[arXiv:1908.02921 [gr-qc]].
%\bibitem{Izumi:2021hlx}
K.~Izumi, Y.~Tomikawa, T.~Shiromizu, and H.~Yoshino,
%``Area bound for surfaces in generic gravitational field,''
PTEP {\bf 2021}, 083E02 (2021).
%[arXiv:2101.03860 [gr-qc]].


\bibitem{Shaikh:2018oul} 
  R.~Shaikh, P.~Banerjee, S.~Paul, and T.~Sarkar,
  %``A novel gravitational lensing feature by wormholes,''
  Phys.\ Lett.\ B {\bf 789}, 270 (2019)
  Erratum: [Phys.\ Lett.\ B {\bf 791}, 422 (2019)].
%  [arXiv:1811.08245 [gr-qc]].

\bibitem{Godani:2021atr}
N.~Godani and G.~C.~Samanta,
%``Gravitational lensing effect in traversable wormholes,''
Annals Phys. {\bf 429}, 168460 (2021).
%[arXiv:2105.08517 [gr-qc]].

\bibitem{Gan:2021pwu}
Q.~Gan, P.~Wang, H.~Wu, and H.~Yang,
%``Photon spheres and spherical accretion image of a hairy black hole,''
Phys. Rev. D {\bf 104}, 024003 (2021).
%[arXiv:2104.08703 [gr-qc]].

\bibitem{Gan:2021xdl}
Q.~Gan, P.~Wang, H.~Wu, and H.~Yang,
%``Photon ring and observational appearance of a hairy black hole,''
Phys. Rev. D {\bf 104}, 044049 (2021). 
%[arXiv:2105.11770 [gr-qc]].

\bibitem{Peng:2021osd}
J.~Peng, M.~Guo, and X.~H.~Feng,
%``Observational signature and additional photon rings of an asymmetric thin-shell wormhole,''
Phys. Rev. D {\bf 104}, 124010 (2021).
%[arXiv:2102.05488 [gr-qc]].

\bibitem{Guerrero:2021pxt}
M.~Guerrero, G.~J.~Olmo, and D.~Rubiera-Garcia,
%``Double shadows of reflection-asymmetric wormholes supported by positive energy thin-shells,''
JCAP {\bf 04}, 066 (2021)
%[arXiv:2102.00840 [gr-qc]].

\bibitem{Wang:2020emr}
X.~Wang, P.~C.~Li, C.~Y.~Zhang, and M.~Guo,
%``Novel shadows from the asymmetric thin-shell wormhole,''
Phys. Lett. B {\bf 811}, 135930 (2020).
%[arXiv:2007.03327 [gr-qc]].

\bibitem{Wielgus:2020uqz}
M.~Wielgus, J.~Horak, F.~Vincent, and M.~Abramowicz,
%``Reflection-asymmetric wormholes and their double shadows,''
Phys. Rev. D {\bf 102}, 084044 (2020).
%[arXiv:2008.10130 [gr-qc]].

\bibitem{Holz:2002uf}
  D.~E.~Holz and J.~A.~Wheeler,
  %``Retro-machos: PI in the sky?,''
  Astrophys.\ J.\  {\bf 578}, 330 (2002).
  %  [astro-ph/0209039].

\bibitem{DePaolis:2003ad}
  F.~De Paolis, G.~Ingrosso, A.~Geralico, and A.~A.~Nucita,
  %``The Black hole at the galactic center as a possible retro-lens for the S2 orbiting star,''
  Astron.\ Astrophys.\  {\bf 409}, 809 (2003).
  %  [astro-ph/0307493].

\bibitem{DePaolis:2004xe}
  F.~De Paolis, A.~Geralico, G.~Ingrosso, A.~A.~Nucita, and A.~Qadir,
  %``Kerr black holes as retro-MACHOs,''
  Astron.\ Astrophys.\  {\bf 415}, 1 (2004).
  %  [astro-ph/0401384].

\bibitem{Abdujabbarov:2017pfw}
A.~Abdujabbarov, B.~Ahmedov, N.~Dadhich and F.~Atamurotov,
%``Optical properties of a braneworld black hole: Gravitational lensing and retrolensing,''
Phys. Rev. D {\bf 96}, 084017 (2017).

\bibitem{Eiroa:2003jf} 
  E.~F.~Eiroa and D.~F.~Torres,
  %``Strong field limit analysis of gravitational retro lensing,''  
  Phys.\ Rev.\ D {\bf 69}, 063004 (2004).
  %[gr-qc/0311013].  

\bibitem{Bozza:2004kq} 
  V.~Bozza and L.~Mancini,
  %``Gravitational lensing by black holes: A Comprehensive treatment and the case of the star S2,''
  Astrophys.\ J.\  {\bf 611}, 1045 (2004).
%  [astro-ph/0404526].

\bibitem{Tsukamoto:2016oca}
N.~Tsukamoto and Y.~Gong,
%``Retrolensing by a charged black hole,''
Phys. Rev. D {\bf 95}, 064034 (2017).
%[arXiv:1612.08250 [gr-qc]].

\bibitem{Tsukamoto:2017edq} 
  N.~Tsukamoto,
  %``Retrolensing by a wormhole at deflection angles $\pi$ and $3\pi$,''
  Phys.\ Rev.\ D {\bf 95}, 084021 (2017).
%  [arXiv:1701.09169 [gr-qc]].

\bibitem{ZamanBabar:2021zuk}
G.~Zaman Babar, F.~Atamurotov, and A.~Zaman Babar,
%``Retrolensing by a spherically symmetric naked singularity,''
[arXiv:2104.01340 [gr-qc]].

\bibitem{Tsukamoto:2021lpm}
N.~Tsukamoto,
%``Retrolensing by light rays slightly inside and outside of a photon sphere around a Reissner-Nordstr\"om naked singularity,''
Phys. Rev. D {\bf 105} 024009 (2022). 
%[arXiv:2109.00495 [gr-qc]].

\bibitem{Visser_1995}
M. Visser,
\textit{Lorentzian Wormholes: From Einstein to Hawking} (American Institute of Physics, Woodbury, NY, 1995).

\bibitem{Morris_Thorne_1988}
M. S. Morris and K. S. Thorne,
Am. J. Phys. {\bf 56}, 395 (1988).

\bibitem{Ohgami:2015nra}
T.~Ohgami and N.~Sakai,
%``Wormhole shadows,''
Phys.\ Rev.\ D {\bf 91}, 124020 (2015).

\bibitem{Paul:2019trt}
S.~Paul, R.~Shaikh, P.~Banerjee, and T.~Sarkar,
%``Observational signatures of wormholes with thin accretion disks,''
JCAP {\bf 03}, 055 (2020).
%[arXiv:1911.05525 [gr-qc]].

\bibitem{Kasuya:2021cpk}
S.~Kasuya and M.~Kobayashi,
%``Throat effects on shadows of Kerr-like wormholes,''
Phys. Rev. D {\bf 103}, 104050 (2021).
%[arXiv:2103.13086 [gr-qc]].

\bibitem{Nambu:2019sqn} 
  Y.~Nambu, S.~Noda, and Y.~Sakai,
  %``Wave Optics in Spacetimes with Compact Gravitating Object,''
  Phys.\ Rev.\ D {\bf 100}, 064037 (2019).
%  [arXiv:1905.01793 [gr-qc]].

\bibitem{Bronnikov:2005gm}
K.~A.~Bronnikov and J.~C.~Fabris,
%``Regular phantom black holes,''
Phys. Rev. Lett. {\bf 96} 251101 (2006).
%[arXiv:gr-qc/0511109 [gr-qc]].

\bibitem{Bronnikov:2006fu}
K.~A.~Bronnikov, V.~N.~Melnikov, and H.~Dehnen,
%``Regular black holes and black universes,''
Gen. Rel. Grav. {\bf 39} 973 (2007).
%[arXiv:gr-qc/0611022 [gr-qc]].

\bibitem{Damour:2007ap} 
  T.~Damour and S.~N.~Solodukhin,
  %``Wormholes as black hole foils,''
  Phys.\ Rev.\ D {\bf 76}, 024016 (2007)
%  [arXiv:0704.2667 [gr-qc]].

\bibitem{Simpson:2018tsi}
A.~Simpson and M.~Visser,
%``Black-bounce to traversable wormhole,''
JCAP {\bf 02}, 042 (2019).
%[arXiv:1812.07114 [gr-qc]].

\bibitem{Nascimento:2020ime}
J.~R.~Nascimento, A.~Y.~Petrov, P.~J.~Porfirio, and A.~R.~Soares,
%``Gravitational lensing in a black-bounce traversable wormhole spacetime,''
Phys. Rev. D {\bf 102}, 044021 (2020).
%[arXiv:2005.13096 [gr-qc]].

\bibitem{Ovgun:2020yuv}
A.~\"Ovg\"un,
%``Weak Deflection Angle of Black-bounce Traversable Wormholes Using Gauss-Bonnet Theorem in the Dark Matter Medium,''
Turk. J. Phys. {\bf 44}, 465 (2020).
%[arXiv:2011.04423 [gr-qc]].

\bibitem{Tsukamoto:2020bjm}
N.~Tsukamoto,
%``Gravitational lensing in the Simpson-Visser black-bounce spacetime in a strong deflection limit,''
Phys. Rev. D {\bf 103}, 024033 (2021).
%[arXiv:2011.03932 [gr-qc]].

\bibitem{Cheng:2021hoc}
X.~T.~Cheng and Y.~Xie,
%``Probing a black-bounce, traversable wormhole with weak deflection gravitational lensing,''
Phys. Rev. D {\bf 103}, 064040 (2021).

\bibitem{Lima:2021las}
H.~C.~D.~Lima, Junior., L.~C.~B.~Crispino, P.~Cunha, V.P., and C.~A.~R.~Herdeiro,
%``Can different black holes cast the same shadow?,''
Phys. Rev. D {\bf 103}, 084040 (2021).
%[arXiv:2102.07034 [gr-qc]].

\bibitem{Bronnikov:2021liv}
K.~A.~Bronnikov, R.~A.~Konoplya, and T.~D.~Pappas,
%``General parametrization of wormhole spacetimes and its application to shadows and quasinormal modes,''
Phys. Rev. D {\bf 103}, 124062 (2021).
%[arXiv:2102.10679 [gr-qc]].

\bibitem{Guerrero:2021ues}
M.~Guerrero, G.~J.~Olmo, D.~Rubiera-Garcia and D.~S.~C.~G\'omez,
%``Shadows and optical appearance of black bounces illuminated by a thin accretion disk,''
JCAP {\bf 08}, 036 (2021).
%[arXiv:2105.15073 [gr-qc]].

\bibitem{Bambhaniya:2021ugr}
P.~Bambhaniya, S.~K, K.~Jusufi and P.~S.~Joshi,
%``Thin accretion disk in the Simpson-Visser black-bounce and wormhole spacetimes,''
Phys. Rev. D {\bf 105}, 023021 (2022).
%[arXiv:2109.15054 [gr-qc]].

\bibitem{Schee:2021pdt}
J.~Schee and Z.~Stuchl\'\i{}k,
%``Appearance of Keplerian discs orbiting on both sides of reflection-symmetric wormholes,''
JCAP {\bf 01}, 054 (2022).
%[arXiv:2111.00750 [gr-qc]].

\bibitem{DellaMonica:2021fdr}
R.~Della Monica and I.~de Martino,
%``Unveiling the nature of SgrA* with the geodesic motion of S-stars,''
JCAP {\bf 03}, 007 (2022).
%[arXiv:2112.01888 [astro-ph.GA]].

\bibitem{Stuchlik:2021tcn}
Z.~Stuchl\'\i{}k and J.~Vrba,
%``Epicyclic Oscillations around Simpson\textendash{}Visser Regular Black Holes and Wormholes,''
Universe {\bf 7}, 279 (2021).
%[arXiv:2108.09562 [gr-qc]].

\bibitem{Bronnikov:2021uta}
K.~A.~Bronnikov and R.~K.~Walia,
%``Field sources for Simpson-Visser spacetimes,''
Phys. Rev. D {\bf 105}, 044039 (2022).
%[arXiv:2112.13198 [gr-qc]].

\bibitem{Mazza:2021rgq}
J.~Mazza, E.~Franzin, and S.~Liberati,
%``A novel family of rotating black hole mimickers,''
JCAP {\bf 04}, 082 (2021).
%[arXiv:2102.01105 [gr-qc]].

\bibitem{Shaikh:2021yux}
R.~Shaikh, K.~Pal, K.~Pal and T.~Sarkar,
%``Constraining alternatives to the Kerr black hole,''
Mon. Not. Roy. Astron. Soc. {\bf 506}, 1229 (2021).
%[arXiv:2102.04299 [gr-qc]].

\bibitem{Islam:2021ful}
S.~U.~Islam, J.~Kumar and S.~G.~Ghosh,
%``Strong gravitational lensing by rotating Simpson-Visser black holes,''
JCAP {\bf 10}, 013 (2021).
%[arXiv:2104.00696 [gr-qc]].

\bibitem{Jiang:2021ajk}
X.~Jiang, P.~Wang, H.~Yang and H.~Wu,
%``Testing Kerr black hole mimickers with quasi-periodic oscillations from GRO J1655-40,''
Eur. Phys. J. C {\bf 81}, 1043 (2021).
%[arXiv:2107.10758 [gr-qc]].

\bibitem{Huang:2019arj}
H.~Huang and J.~Yang,
%``Charged Ellis Wormhole and Black Bounce,''
Phys. Rev. D {\bf 100}, 124063 (2019).
%[arXiv:1909.04603 [gr-qc]].

\bibitem{Lobo:2020ffi}
F.~S.~N.~Lobo, M.~E.~Rodrigues, M.~V.~d.~S.~Silva, A.~Simpson, and M.~Visser,
%``Novel black-bounce spacetimes: wormholes, regularity, energy conditions, and causal structure,''
Phys. Rev. D {\bf 103}, 084052 (2021)
%[arXiv:2009.12057 [gr-qc]].

\bibitem{Franzin:2021vnj}
E.~Franzin, S.~Liberati, J.~Mazza, A.~Simpson, and M.~Visser,
%``Charged black-bounce spacetimes,''
JCAP {\bf 07}, 036 (2021).
%[arXiv:2104.11376 [gr-qc]].

\bibitem{Xu:2021lff}
Z.~Xu and M.~Tang,
%``Rotating spacetime: black-bounces and quantum deformed black hole,''
Eur. Phys. J. C \textbf{81} (2021) no.10, 863
%[arXiv:2109.13813 [gr-qc]].

\bibitem{Chatzifotis:2021hpg}
N.~Chatzifotis, E.~Papantonopoulos and C.~Vlachos,
%``Disformal transition of a black hole to a wormhole in scalar-tensor Horndeski theory,''
Phys. Rev. D {\bf 105}, 064025 (2022).
%[arXiv:2111.08773 [gr-qc]].

\bibitem{Guo:2021wid}
Y.~Guo and Y.~G.~Miao,
%``Charged black-bounce spacetimes: Photon rings, shadows and observational appearances,''
[arXiv:2112.01747 [gr-qc]].

\bibitem{Barrientos:2022avi}
J.~Barrientos, A.~Cisterna, N.~Mora, and A.~Vigan\`o,
%``(A)dS Taub-NUT and exact black bounces with scalar hair,''
[arXiv:2202.06706 [hep-th]].

\bibitem{Ellis_1973} 
H. G. Ellis, 
J. Math. Phys. {\bf 14}, 104 (1973); 
%\bibitem{Bronnikov_1973}
K. A. Bronnikov, 
Acta Phys. Pol. B {\bf 4}, 251 (1973);
%\bibitem{Martinez:2020hjm}
C.~Martinez and M.~Nozawa,
%``Static spacetimes haunted by a phantom scalar field. I. Classification and global structure in the massless case,''
Phys. Rev. D {\bf 103}, 024003 (2021).
%[arXiv:2010.05183 [gr-qc]].

\bibitem{Chetouani_Clement_1984}
L. Chetouani and G. Cl\'{e}ment,
Gen. Relativ. Gravit. {\bf 16}, 111 (1984);
%\bibitem{Nakajima_Asada_2012} 
K. Nakajima and H. Asada, 
Phys. Rev. D {\bf 85}, 107501 (2012);
%\bibitem{Tsukamoto_Harada_Yajima_2012} 
N. Tsukamoto, T. Harada, and K. Yajima,
Phys. Rev. D {\bf 86}, 104062 (2012);
F. Abe, 
Astrophys. J. {\bf 725}, 787 (2010);
%\bibitem{Toki_Kitamura_Asada_Abe_2011} 
Y. Toki, T. Kitamura, H. Asada, and F. Abe, 
Astrophys. J. {\bf 740}, 121 (2011); 
%\bibitem{Tsukamoto_Harada_2013} 
N. Tsukamoto and T. Harada,
Phys. Rev. D {\bf 87}, 024024 (2013); 
%\bibitem{Takahashi_Asada_2013} 
R. Takahashi and H. Asada,
Astrophys. J. {\bf 768}, L16 (2013);
N.~Tsukamoto and Y.~Gong,
%``Extended source effect on microlensing light curves by an Ellis wormhole,''
Phys. Rev. D {\bf 97}, 084051 (2018).
%[arXiv:1711.04560 [gr-qc]].

\bibitem{Tsukamoto:2021caq}
N.~Tsukamoto,
%``Gravitational lensing by two photon spheres in a black-bounce spacetime in strong deflection limits,''
Phys. Rev. D {\bf 104} 064022 (2021).
%[arXiv:2105.14336 [gr-qc]].

\bibitem{Guerrero:2022qkh}
M.~Guerrero, G.~J.~Olmo, D.~Rubiera-Garcia, and D.~S.~C.~G\'omez,
%``Light ring images of double photon spheres in black hole and wormhole space-times,''
[arXiv:2202.03809 [gr-qc]].

\bibitem{Eiroa:2002mk} 
  E.~F.~Eiroa, G.~E.~Romero, and D.~F.~Torres,
  %``Reissner-Nordstrom black hole lensing,''
  Phys.\ Rev.\ D {\bf 66}, 024010 (2002).
%  [gr-qc/0203049].

\bibitem{Bozza:2005tg} 
  V.~Bozza, F.~De Luca, G.~Scarpetta, and M.~Sereno,
  %``Analytic Kerr black hole lensing for equatorial observers in the strong deflection limit,''
  Phys.\ Rev.\ D {\bf 72}, 083003 (2005).
  %[gr-qc/0507137].

\bibitem{Bozza:2006nm} 
  V.~Bozza, F.~De Luca, and G.~Scarpetta,
  %``Kerr black hole lensing for generic observers in the strong deflection limit,''
  Phys.\ Rev.\ D {\bf 74}, 063001 (2006).
%  [gr-qc/0604093].

\bibitem{Iyer:2006cn}
  S.~V.~Iyer and A.~O.~Petters,
  %``Light's bending angle due to black holes: From the photon sphere to infinity,''
  Gen.\ Rel.\ Grav.\  {\bf 39}, 1563 (2007).
%  [gr-qc/0611086].

\bibitem{Bozza:2007gt} 
V.~Bozza and G.~Scarpetta,
%``Strong deflection limit of black hole gravitational lensing with arbitrary source distances,''
Phys.\ Rev.\ D {\bf 76}, 083008 (2007).
%[arXiv:0705.0246 [gr-qc]].

\bibitem{Tsukamoto:2016qro} 
  N.~Tsukamoto,
  %``Strong deflection limit analysis and gravitational lensing of an Ellis wormhole,''
  Phys.\ Rev.\ D {\bf 94}, 124001 (2016).
%  [arXiv:1607.07022 [gr-qc]].

\bibitem{Ishihara:2016sfv} 
  A.~Ishihara, Y.~Suzuki, T.~Ono, and H.~Asada,
  %``Finite-distance corrections to the gravitational bending angle of light in the strong deflection limit,''
  Phys.\ Rev.\ D {\bf 95}, 044017 (2017).
%  [arXiv:1612.04044 [gr-qc]].

\bibitem{Tsukamoto:2016jzh} 
  N.~Tsukamoto,
  %``Deflection angle in the strong deflection limit in a general asymptotically flat, static, spherically symmetric spacetime,''
  Phys.\ Rev.\ D {\bf 95}, 064035 (2017).
%  [arXiv:1612.08251 [gr-qc]].

\bibitem{Aldi:2016ntn}
G.~F.~Aldi and V.~Bozza,
%``Relativistic iron lines in accretion disks: the contribution of higher order images in the strong deflection limit,''
JCAP {\bf 02}, 033 (2017).
%[arXiv:1607.05365 [astro-ph.HE]].

\bibitem{Hsieh:2021scb}
T.~Hsieh, D.~S.~Lee, and C.~Y.~Lin,
%``Strong gravitational lensing by Kerr and Kerr-Newman black holes,''
Phys.\ Rev.\ D {\bf 103}, 104063 (2021).
%[arXiv:2101.09008 [gr-qc]].

\bibitem{Takizawa:2021gdp}
K.~Takizawa and H.~Asada,
%``Iterative solutions for the gravitational lens equation in the strong deflection limit,''
Phys. Rev. D {\bf 103}, 104039 (2021).
%[arXiv:2103.10649 [gr-qc]].

\bibitem{Bisnovatyi-Kogan:2022ujt}
G.~S.~Bisnovatyi-Kogan and O.~Y.~Tsupko,
%``Analytical study of higher-order ring images of accretion disk around black hole,''
[arXiv:2201.01716 [gr-qc]].

\bibitem{Bozza:2008ev}
V.~Bozza,
%``A Comparison of approximate gravitational lens equations and a proposal for an improved new one,''
Phys. Rev. D {\bf 78}, 103005 (2008).
%[arXiv:0807.3872 [gr-qc]].

\bibitem{Witt:1994}
H.~J.~Witt and S.~Mao, ApJ, {\bf 430}, 505 (1994).

\bibitem{Nemiroff:1994uz}
  R.~J.~Nemiroff and W.~A.~D.~T.~Wickramasinghe,
  %``Finite source sizes and the information content of MACHO type lens search light curves,''
  Astrophys.\ J.\  {\bf 424}, L21 (1994).

\bibitem{Alcock:1997fi}
  C.~Alcock {\it et al.} [MACHO and GMAN Collaborations],
  %``MACHO alert 95-30: First real time observation of extended source effects in gravitational microlensing,''
  Astrophys.\ J.\  {\bf 491}, 436 (1997).

\bibitem{Tsukamoto}
N.~Tsukamoto, in preparation.


\end{thebibliography}
\end{document}